\documentclass[aps,prb,reprint,superscriptaddress,floatfix]{revtex4-2}

\newcommand{\strain}    {\ensuremath{\epsilon}}
\usepackage{graphicx}
\usepackage{amssymb}
\usepackage{amsmath}
\newcommand{\half}{\ensuremath{\frac{1}{2}}}
\newcommand{\halfl}{\ensuremath{{\scriptstyle \frac{1}{2}}}}

\newcommand{\un}   [1]{\ensuremath{\,\mathrm{#1}}}

\newcommand{\intd}    {\ensuremath{\,\mathrm{d }}}

\newcommand{\pder} [2]{\ensuremath{\frac{\partial #1}{\partial #2}}}
\newcommand{\pderl}[2]{\ensuremath{\partial #1/\partial #2}}

\newcommand{\mySiN}{Si\textsubscript{3}N\textsubscript{4}~}

\ifdefined\eqref
\else
    \newcommand{\eqref}[1]{(\ref{#1})}
\fi

\begin{document}


\title{Relaxation and dynamics of high-stress pre-displaced string resonators}

 
\author{Xiong Yao}
\affiliation{Department of Physics, Technical University Munich, Garching, Germany} 

\author{David Hoch}
\affiliation{Department of Physics, Technical University Munich, Garching, Germany} 
\affiliation{Munich Center for Quantum Science and Technology (MCQST), Munich, Germany} 
\affiliation{Institute for Advanced Study, Technical University Munich, Garching, Germany}

\author{Menno Poot}
\affiliation{Department of Physics, Technical University Munich, Garching, Germany} 
\affiliation{Munich Center for Quantum Science and Technology (MCQST), Munich, Germany} 
\affiliation{Institute for Advanced Study, Technical University Munich, Garching, Germany}
\email{menno.poot@tum.de} 

\date{\today}

\begin{abstract}
Pre-displaced micromechanical resonators made from high-stress material give rise to new rich static and dynamic behavior. Here, an analytical model is presented to describe the mechanics of such pre-displaced resonators. The bending and tension energies are derived and a modified Euler-Bernoulli equation is obtained by applying the least action principle. By projecting the model onto a cosine shape, the energy landscape is visualized, and the pre-displacement dependence of stress and frequency is studied semi-analytically. The analysis is extended with finite-element simulations, including the mode shape, the role of overhang, and the stress distribution.
\end{abstract}

\maketitle

\section{Introduction}
Nowadays, micro and nanomechanical resonators are widely used for a large variety of applications, ranging from sensing and detection \cite{westerveld_natphot_ultrasound_sensor, calleja_rscnanoscale_nanomech_sensing, Waggoner_rsc_micronanomech_sensing}, 
optical and microwave quantum transducers in hybrid opto-electro-mechanical systems \cite{lauk2020perspectives}, to fundamental experiments in the quantum regime \cite{oconnell_nature_quantum_piezo_resonator,chan2011laser,fiaschi_natphot_optomechanical_quantum_teleportation}. These resonators are often made out of pre-strained films, like \mySiN, as such materials provide very high quality factors \cite{poot_NJP_Yfeedback,Hoej_NatureCommunications_trampoline_inverse_design,Unterreithmeier_PRL_damping,Norte_PhysicalReviewLetters_SiN_yield_stress}, enabling e.g. high detection efficiencies in the case of sensors \cite{Waggoner_rsc_micronanomech_sensing} and long coherence times in mechanical quantum storage \cite{Heinrich_NatureNanotechnology_quantum_coherence}. For all these experiments, it is imperative to engineer the geometry of the resonator \cite{eichenfield_nature_optomechanical_crystals, Hoej_NatureCommunications_trampoline_inverse_design, poot_apl_phaseshifter, Bagheri_PRL_cavity_sync, cole_natcomm_phonon_tunnelling, Fong_NanoLetters_phonon_coupling, Bereyhi_arXiv_ultrahighQ} including the stress \cite{Ghadimi_Science_strain_engineering, Beccari2022_nature_10B_quality}, to reach the best performance. Our recent work \cite{Hoch_Sbeam} experimentally demonstrates the geometrical tuning of the stress of resonators made out of pre-strained SiN films. This not only provides a new approach for a systematic study of dissipation dilution \cite{Unterreithmeier_PRL_damping, Schmid_PhysRevB_damping_model} but also raises interesting questions, like their potential to buckle and the influence of the geometry on the eigenmodes. A thorough study of the statics and dynamics of these structures is, however, still lacking.

Here, we provide a detailed analysis of the mechanics of pre-displaced beam resonators using analytical methods which are supported by finite-element simulations. In Sec.~\ref{sec:relaxation} the stress relaxation of straight and pre-displaced high-stress beams is studied, as well as their potential to buckle. In Sec. \ref{sec:energy}, expressions for the bending and tension energy stored in the beam are derived. Based on this, their equations of motion - i.e. modified Euler-Bernoulli equations - are obtained in Sec. \ref{sec:EoM}. For a deeper understanding, the model is projected onto the mode shape resembling the fundamental mode (Sec. \ref{sec:projection}). Finally, in Sec. \ref{sec:FEM}, finite elements simulations are used to validate the results and to study the role of the overhang, the mode shape, and the distribution of the stress throughout the beams.

\section{Relaxation} \label{sec:relaxation}
\begin{figure}[!tbhp]
  \includegraphics[width=\columnwidth]{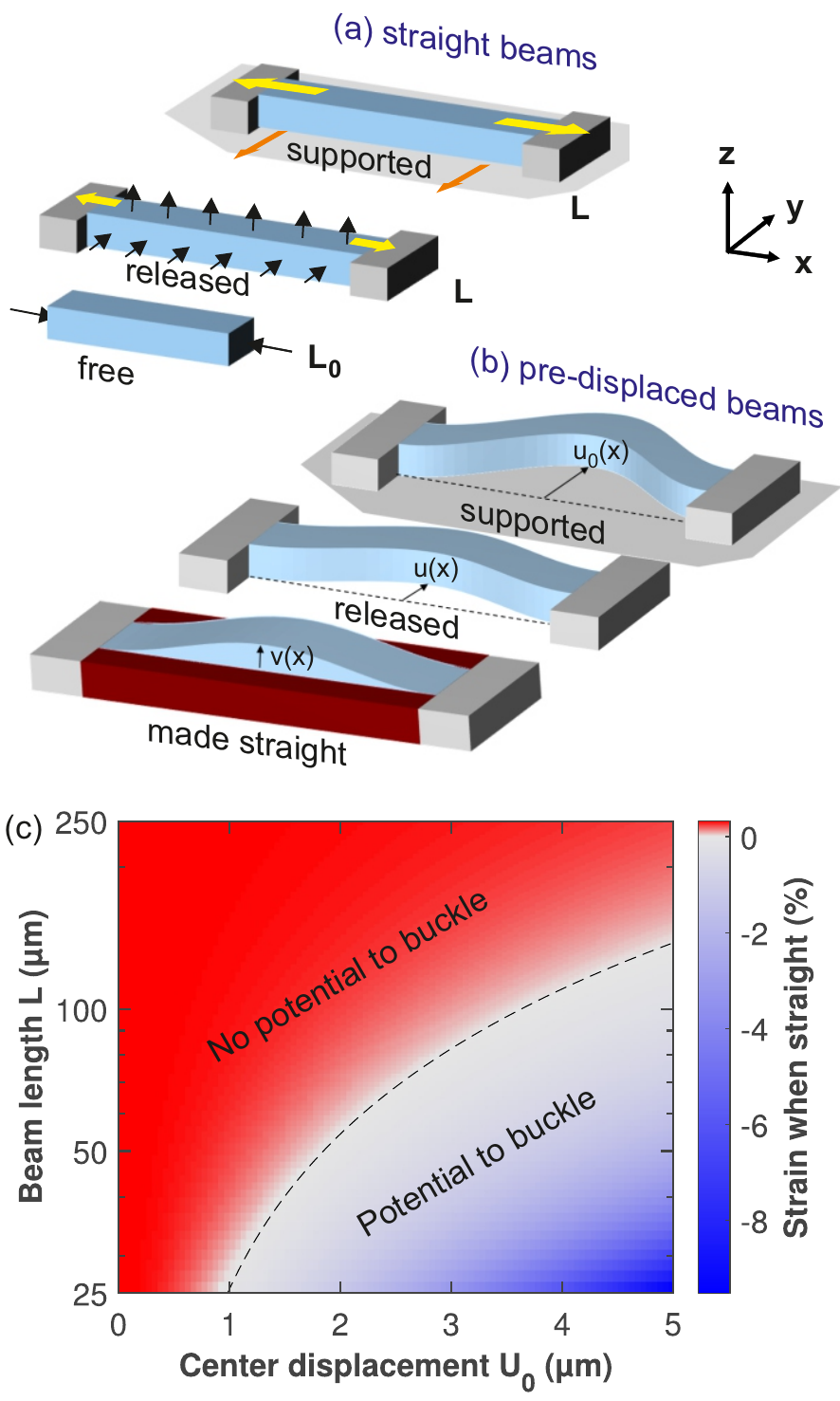}
  \caption{
  (a) Illustration of the relaxation of straight strings. In the supported case (top), there is both a large stress in the x direction (yellow) and y direction (orange). After release (middle), the stress in the y direction has relaxed, resulting in shrinking in the y direction (in-plane, black arrows). For typical materials with a positive Poisson ration, this gives rise to an expansion in the z direction (out-of-plane, black arrows) and a reduction in the stress in the x direction (yellow). The length is still the separation between the clamping points. If the beam would be completely free (bottom), its length would be $L_0$ and there are no stresses anymore. Here, for clarity, the transverse deformations are not shown. 
  (b) Relaxation of pre-displaced beams. When supported (top), the beams have designed pre-displacement profile $u_0(x)$ and after release (middle) this relaxes to $u(x)$. If the beams are forced to be be straight (bottom), e.g. by pushing with strong structures from the sides (red), an out-of-plane buckling displacement $v(x)$ can result.
  (c) Colorplot of the resulting strain if the beams were made straight [as in the bottom of panel (b)]. Compressive (tensile) strains are indicated in blue (red) and the dashed line indicates the critical strain for out-of-plane buckling $\strain_{c,z}$. Note that $\strain_c$ is almost indistinguishable from $\strain = 0$. For a complete list of parameters, see Appendix.
  }
  \label{fig:relaxation}
\end{figure}

\subsection{Relaxation of straight strings} \label{ssec:relax_straight}
Nanomechanical strings are typically made by etching a thin film, most notably silicon nitride, with a large, uniform tensile stress, followed by their release from the supporting substrate. The films are typically thin (thickness $h \ll$ length $L$) so that the out-of-plane components (i.e., $z.$; see Fig.~\ref{fig:relaxation}(a) for the coordinate system) of the stress tensor vanish ($\sigma_{iz} = \sigma_{zi} =0$ for $i\in\{x,y,z\}$), and only in-plane tensile forces remain. 
For an isotropic material, there are no shear stresses $\sigma_{xy} = \sigma_{yx} = 0$ and only the components $\sigma_{xx} = \sigma_{yy} \equiv \sigma_\mathrm{film}$ remain.
The film stress $\sigma_\mathrm{film}$ is a property of the growth process and can be controlled using e.g. the stoichiometry of the material. It is the amount of stress that remains present for two-dimensional resonators that are clamped on all sides, such as membranes \cite{hoch_MM_mode_mapping}, or when narrow (width $w \ll L$) and straight structures are patterned but still held by the supporting substrate as illustrated in the top panel of Fig. \ref{fig:relaxation}(a). After release, the forces in the $y$ direction cannot be sustained and $\sigma_{yy} \approx 0$ [Fig. \ref{fig:relaxation}(a) center]. In this case, $\sigma_{xx}$ is, thus, the only remaining stress component, and we define its value as $\sigma_0$. Since the straight beam is clamped at the ends, the length before and after the release remains the same [see Fig. \ref{fig:relaxation}(a)] and thus the xx-component of the strain tensor, $\strain_{xx}$ is identical before and after the release. For an isotropic linear-elastic material with Young's modulus $E$ and Poison ratio $\nu$, this longitudinal strain is given by $\strain_{xx} = (\sigma_{xx} - \nu \sigma_{yy})/E$, which is $(1-\nu)\sigma_\mathrm{film}/E$ before and $\sigma_0/E$ after the release. Since these two are equal, the remaining xx-component of the stress tensor $\sigma_0$ can be determined: $\sigma_0 = (1-\nu)\sigma_\mathrm{film}$. For our silicon nitride films with $\sigma_\mathrm{film} = 1050 \un{MPa}$ and $\nu = 0.23$, this yields $\sigma_0 = 809 \un{MPa}$ (see Appendix), corresponding to a pre-strain $\strain_{xx} = \sigma_0/E = 0.31\%$. Note that the cross sectional area $A = hw$ can always be used to convert back and forth between the stress $\sigma_{xx}$ and tension $T = \sigma_{xx}A$.

Finally, when the clamping at $x=0$ and $x=L$ would be removed [Fig. \ref{fig:relaxation}(a) bottom], the beam would relax completely so that also $\sigma_{xx} = 0$, and the beam would attain a length $L_0 = (1-\strain_{xx})L$. To put this into perspective: a $100 \un{\mu m}$ long beam would shrink by $L-L_0 = 310 \un{nm}$ when freed.

\subsection{Relaxation of pre-displaced strings} \label{ssec:relax_pre}
So far, the discussion of relaxation of high-stress beams has focused on straight strings.
Figure \ref{fig:relaxation}(b) shows a schematic of our pre-displaced strings \cite{Hoch_Sbeam}. Here, the beam is made with a center line that is not straight, but has a x-dependent displacement in the y-direction $u_0(x)$. The length of the beam $\ell$ is thus longer than then distance between the clamping points $L$; the exact length depends both on $L$ and $u_0(x)$ as will be shown in the next Section. Before release, the beam is supported [Fig. \ref{fig:relaxation}(b) top] and the stress is again $\sigma_\mathrm{film}$. Now, upon release both transverse relaxation, as well as  straightening of its shape will occur. Both will happen together, but for the understanding it is good to imagine this as two separate steps: one where the transverse stress relaxes like was shown in Fig. \ref{fig:relaxation}(a), but  still retaining the pre-displacement, and a second one where the pre-displacement relaxes too. After the first step, the relaxed stress would also be $\sigma_0 = (1-\nu)\sigma_\mathrm{film}$, just like for a straight beam, but now along the direction of the center line, which can be locally under an angle with the x-axis. There will thus be an uncompensated y-component in the tension along the string, which will cause the straightening changing the profile to $u(x)$ [Fig. \ref{fig:relaxation}(b) middle]. As we will detail in the following sections, how much the string straightens depends on the competition between tension and bending energy, and if there is potential to buckle.

\subsection{Potential to buckle} \label{ssec:buckling}
Buckling is the sudden deformation of a structure under a compressive load, which can lead to out-of-plane deformations \cite{etaki_natphys_squid, erbil_PRL_buckling_MEMS, nayfeh_buckled}. In engineering, buckling may result in catastrophic failure of structures, but in micromechanics buckling can also be harnessed to implement a variety of functions in micromechanical devices, e.g. for information storage \cite{charlot_JMM_buckling_memory, bagheri_natnano_high_amplitude} or to control propagation of waves \cite{kim_NL_buckling_SiN_drum_array}. As explained above, when pre-displaced strings relax towards the line connecting the clamping points, their curve length shortens and the tensile stress decreases. When the string would be (actively made) straight between the clamping points [see Fig. \ref{fig:relaxation}(b) bottom], the tension can become negative, and out-of-plane deformations - cf. buckling in the z-direction - may be energetically favorable compared to in-plane deformations for $h < w$. In this Section we explore if this situation can occur, or not.

The curve length $\ell$ of the resonator depends on the in- and out-of-plane displacement profiles, $u(x)$ and $v(x)$, respectively, through the functional
\begin{equation}
    \ell[u,v]  = \int_0^L \left(1+\left(\pder{u}{x}\right)^2 + \left(\pder{v}{x}\right)^2\right)^{1/2} \intd x.     \label{eq:ell-full}
\end{equation}
Throughout this paper, it is assumed that the displacements are not large, e.g. $u_0(x) \ll L$ and $\pderl{u_0}{x} \ll 1$ and likewise for $u$ and $v$. In this approximation, the length becomes
\begin{equation}
    \ell[u,v]  \approx L  + \half \int_0^L \left(\pder{u}{x}\right)^2 + \left(\pder{v}{x}\right)^2 \intd x.
    \label{eq:ell}
\end{equation}
Now, if the string would be completely straight (cf. $u(x) = 0$ and $v(x) = 0$), then the curve length should equal the distance between the clamping points: $\ell[0,0] = L$, which is not necessarily equal to the length that the straightened string would have if is was not clamped, $L_0$, thus resulting in a strain $\strain_{xx} = (L-L_0)/L$ that can be positive or negative, depending on whether $L_0$ is smaller or larger than $L$. For $L_0<L$ there is still tensile strain and no buckling occurs. However, if $L_0>L$ there is a compressive load exerted by the clamping points. Still, out-of-plane buckling may only occur when that compression is large enough, i.e., when the critical strain $\strain_{c,z} = -4\pi^2h^2/\{12 L^2 (1-\nu^2)\} < 0$  \cite{poot_physrep_quantum_regime} is exceeded: $(L-L_0)/L < \strain_{c,z}(L)$. 

Figure~\ref{fig:relaxation}(c) shows a color plot of the strain when beams with varying length $L$ and varying pre-displacements $u_0(x) = \halfl U_0 [1-\cos(2\pi x/L)]$ would be made straight, as well as the line where the critical strain $\strain_{c,z}$ is reached (dashed line). Long beams with small initial displacements (upper left corner) still have tensile strain when made straight, but short beams with a large displacement (lower right corner) would have compressive strain that can exceed the critical strain. Even though in Fig. \ref{fig:relaxation}(c) there was a large tensile stress present before release, this shows that there are still beams that have the potential to buckle, and, thus, also the out-of-plane displacement $v$ should be taken into account in the analysis.

\section{Bending and tension energy} \label{sec:energy}
To understand the statics and dynamics of the pre-displaced strings, the potential energy that is stored in both the bending and stretching of the beam is needed. In this section, first the bending energy $E_B$ is calculated, followed by the tension energy $E_T$.  

\subsection{Bending energy} \label{ssec:EB}
It costs energy to deform a mechanical structure and a part of that is due to bending. For example, when a doubly-clamped beam is displaced downwards, in the middle its bottom surface will be stretched, whereas its top becomes compressed \cite{cleland_nanomechanics, LL_elasticity}. Only the neutral plane does not deform. There, not only the direction of the in-plane displacement $u_x$, but also the displacement-induced stretching force (concretely $\sigma_{xx} - \sigma_0$, see Sec. \ref{ssec:ET}) reverses sign there, so that both in the stretched and in the compressed area, elastic energy is stored. By averaging the work needed over the cross section of the beam, one obtains \cite{cleland_nanomechanics, poot_physrep_quantum_regime, Unterreithmeier_PRL_damping} 
\begin{equation}
    U_{B,z}[v(x)] = \frac{D_z}{2}\int_0^L v''^2(x) \intd x, \label{eq:UBz}
\end{equation}
where $D_z = EI_z/(1-\nu^2)$ is the bending rigidity (also known as the flexural rigidity) and $I_z$ is the second moment of area which equals $wh^3/12$ for a beam with a rectangular cross section displaced in the z-direction \cite{cleland_nanomechanics}. The quotes in Eq. \eqref{eq:UBz} denote derivatives with respect to $x$; the bending energy thus depends on the curvature of the displacement profile, $v'' = \pderl{^2v}{x^2}$, squared.  

In the case of a tensionless beam with a pre-displacement in the y direction $u_0(x)$, the same argument can be used to find the bending energy for the in-plane direction: 
\begin{equation}
    U_{B,y}[u(x)] = \frac{D_y}{2}\int_0^L \big(u''(x)-u''_0(x)\big)^2 \intd x,  
    \label{eq:UBy}
\end{equation}
with $D_y = EI_y/(1-\nu^2)$ and $I_y = w^3h/12$. For $u(x) = u_0(x)$, the bending energy is at its minimum: $U_{B,y}[u_0(x)] = 0$ and the more the beam displaces from its initial shape the more bending energy this costs. The total bending energy $E_B$ is the sum of $U_{B,y}$ and $U_{B,z}$.

\subsection{Tension energy} \label{ssec:ET}
Compared to $E_B$, the tension energy $E_T$ is more subtle to calculate since, in addition to a constant component $T_0$, a part of the tension $T$ depends on the flexural displacements.
First focusing on in-plane displacements $u(x)$ only, one can ask what force distribution $f_y(x)$ generates a particular $u(x)$, given a tension $T$. The static force balance of a string under tension is \cite{poot_physrep_quantum_regime}
\begin{equation}
    -T[u(x)] \times \pder{^2u}{x^2} = f_y(x).   \label{eq:fx}   
\end{equation}
This is typically used to find the displacement for a given force distribution by solving the differential equation, but when $u(x)$ is already specified, the force per unit length that is needed to create that displacement can be obtained from Eq.~\eqref{eq:fx} directly. $f_y(x)$ is thus a functional of the displacement profile, we indicate this with $f_y[u(x)](x)$. Eq.~\eqref{eq:fx} shows that the larger the displacement, the larger the force per unit length has to be.

Physically, $f_y(x)$ originates from the tension $T$ that tries to pull the string back to $u(x)=0$. When incrementing the displacement, work is done against that tension, which is stored as potential energy. By summing the work required to bring the displacement from 0 to $u(x)$, and integrating over the length of the string, one obtains:
\begin{equation}
    E_T[u(x)] = \int_0^L \int_0^{u(x)} f_y[\tilde u(x)](x) \intd \tilde u \intd x. \label{eq:ET1}
\end{equation}
By using Eq.~\eqref{eq:fx} and defining $\tilde u(x) = su(x)$, Eq. \eqref{eq:ET1} can be expressed as
\begin{eqnarray}
    E_T[u(x)] & = & \int_0^L \int_0^1 -T[su(x)] \pder{^2(su)}{x^2}  \intd (su) \intd x  \\ 
    & = & \int_0^1 T[su(x)] s \intd s \int_0^L -\pder{^2u}{x^2} u(x) \intd x. \label{eq:ET2}
\end{eqnarray}
Note, that the second integral in Eq.~\eqref{eq:ET2} only contains the \emph{final} displacement profile $u(x)$ and its curvature, whereas the first integral takes the changing magnitude of the displacement and tension during the process of going from $\tilde u =0$ to $\tilde u = u$ into account through the dummy variable $s$. If the tension would be independent of $u(x)$, i.e., $T[u(x)] = T_0$, then the integral over $s$ would give $\halfl T_0$ \cite{Unterreithmeier_PRL_damping}. With the contribution because of the displacement-induced elongation of Eq.~\eqref{eq:ell} included, the tension is 
\begin{equation}
    T[u(x)]  =  T_0 + \frac{AE}{2L}\int_0^L \left( \pder{u}{x} \right)^2 \intd x. \label{eq:T}
\end{equation}
The second term in Eq.~\eqref{eq:T} gives an additional $s^2$, so that $s^4$ appears in the anti-derivative of the first integrant of Eq.~\eqref{eq:ET2}. The prefactor of that term is $\frac{1}{4}$ instead of the $\halfl$ in front of $T_0$ in the expression from $E_T$. Here, it should be emphasized again that $T_0$ is the tension of the straight beam $T[u(x) = 0]$, and \emph{not} the tension $\sigma_0 A = T[u_0(x)]$ initially present after the transverse stress relaxation, as was detailed in Sec.~\ref{ssec:relax_pre}. Still, after inserting $u(x) = u_0(x)$ into Eq.~\eqref{eq:T} and rearranging, one obtains
\begin{equation}
    T_0 = \sigma_0 A - \frac{AE}{2L}\int_0^L \left( \pder{u_0}{x} \right)^2 \intd x. \label{eq:T0}
\end{equation}
which shows clearly that $T_0$ depends on the pre-displacement $u_0$ and on the initial stress $\sigma_0$. 

After performing partial integration and realizing that the boundary terms are zero for the boundary conditions $u(0) = u(L) = 0$, the tension energy becomes: 
\begin{equation}
    E_T  =  \half\Big\{\halfl T_0 + \halfl T[u(x)]\Big\}\int_0^L  \left( \pder{u}{x} \right)^2 \intd x. \label{eq:ET3}
\end{equation}
It should be noted that the ``effective'' tension appearing between the curly brackets in $E_T$ is neither the initial tension $T_0$, nor the final tension $T[u]$, but $T_\mathrm{eff} = \halfl(T_0 + T) = T_0 + \halfl(T[u(x)] - T_0)$. When $T_\mathrm{eff}$ is positive, $u(x) = 0$, i.e. straight strings, is a minimum of $E_T$. When the effective tension is negative, $u(x) = 0$ corresponds to a maximum in $E_T$ and buckling may occur, as explored in Sec. \ref{ssec:buckling}. From Eq.~\eqref{eq:T} it is clear that this requires $T_0 < 0$. 

When both displacements in the y- ($u$) and z-direction ($v$) are present, the tension is a functional of both profiles, $T[u(x),v(x)]$, and after a similar derivation as done above for $u$ only, the tension energy $E_T$ becomes:
\begin{equation}
    E_T  =  \half T_\mathrm{eff} \int_0^L \left( \pder{u}{x} \right)^2 + \left( \pder{v}{x} \right)^2  \intd x. \label{eq:ET4}
\end{equation}
with $T_\mathrm{eff} = \halfl T_0 + \halfl T[u,v]$.

\section{Equations of motion} \label{sec:EoM}
The equation of motion for $u(x,t)$ and $v(x,t)$ can be obtained using the formalism of Lagrangian mechanics \cite{Wells_Lagrangian_dynamics}. For this, the total potential energy $E_\mathrm{tot} = E_B+E_T$, as derived in the previous Section, as well as the kinetic energy $K$ are needed. The latter is: 
\begin{equation}
    K  =  -\half \rho w h \int_0^L \dot u^2 + \dot v^2  \intd x, \label{eq:K}
\end{equation}
where the dot indicates a derivative w.r.t. time $t$.

To obtain the equations of motion, in short, one inserts $u \rightarrow u+\delta u$ (and likewise for $\delta v$) into the Lagrangian $L = K- E_\mathrm{tot}$ and linearizes in the infinitesimal virtual displacement $\delta u(x,t)$. This results in integrals containing $\delta u$, as well as its time (in $K$) and spatial derivatives (in $E_B$, $E_T$, and $T$). And after performing partial integration, and setting the total change in the Lagrangian $\delta L$ to zero, one obtains an equation with an integral over the beam length containing $\delta u(x,t)$ itself, but no longer its derivatives. Since $\delta L = 0$ should hold for arbitrary $\delta u$, the prefactor of $\delta u(x,t)$ inside the integral should vanish at all locations $x$. This yields to the Euler-Bernoulli equations \cite{cleland_nanomechanics, LL_elasticity} with tension included \cite{nayfeh_buckled, poot_PSSB_modelling_CNT, westra_PRL_coupled, Unterreithmeier_PRL_damping}, but now with the fourth order spatial derivative of $u-u_0$ instead of $u$:
\begin{eqnarray}
\rho A \ddot u & = & -D_y \left(\pder{^4u}{x^4}-\pder{^4u_0}{x^4}\right) + T \pder{^2u}{x^2} + f_y(x,t) \label{eq:EulerBernoulli_u}\\
\rho A \ddot v & = & -D_z \pder{^4v}{x^2} + T \pder{^4v}{x^2} + f_z(x,t). \label{eq:EulerBernoulli_v}
\end{eqnarray}
Here, external forces (per unit length) in the y and z direction ($f_y(x,t)$ and $f_z(x,t)$) have also been included. Note, that the \emph{actual} tension $T[u,v]$ appears again \cite{nayfeh_buckled}, and not the effective tension appearing as prefactor in $E_T$ [see Eq.~\eqref{eq:ET4}]. This is because the virtual work $\delta E_T$ done by the virtual displacement $\delta u$ not only contains the direct change $T_\mathrm{eff} \int_0^L u'\delta u' \intd x$ via the integral of Eq.~\eqref{eq:ET4}, but also the change $\delta T$. This is analogous to the emergence of the ac tension \cite{westra_PRL_coupled} in description of the flexural resonances of carbon nanotubes \cite{poot_PSSB_modelling_CNT} and buckled beams \cite{etaki_natphys_squid}. The fact that only $T$ appears in the equation of motion is  expected, since in a local force balance - which would also lead to Eqs.~\eqref{eq:EulerBernoulli_u}-\eqref{eq:EulerBernoulli_v} - it is irrelevant if the tension is due to $T_0$, due to the elongation, or a combination of the two. 

A dimension analysis \cite{poot_PSSB_modelling_CNT} shows that $TL^2/D_y \propto (\sigma/E) (L/w)^2 \equiv \Sigma_y$ is the parameter that determines the importance of tension over bending rigidity. A resonator with $\Sigma_y \gg 1$ behaves as a string, whereas one with $\Sigma_y \ll 1$ acts as a tensionless beam.
Interestingly, when the ``stringness'' \cite{Hoch_Sbeam} $\Sigma_y \gg 1$, i.e. a resonator where the tension dominates over the bending rigidity, $u_0$ drops out of Eq.~\eqref{eq:EulerBernoulli_u}. Thus, after relaxing, the pre-displaced beams simply behave as strings under tension and the only effect of the pre-displacement $u_0$ will be the geometric tuning of the tension $T$ \cite{Hoch_Sbeam}.
Irrespective of the value of $\Sigma_y$, the $u_0''''$ term in Eq.~\eqref{eq:EulerBernoulli_u} is independent of $u$ and $v$ and, hence, when solving the equations of motion, that term can be viewed as an additional in-plane force per unit length $+D_y u_0''''$ that acts on the beam. Thus, to solve the static displacement and the eigenmodes, one can follow the standard approach of inserting $u(x,y) = u_\mathrm{dc}(x) + u_\mathrm{ac}(x,t)$ and solving for the static profile $u_\mathrm{dc}(x)$ and for the eigenmodes by taking $u_\mathrm{ac}(x,t) = \chi_n(x)\exp(i\omega_n t)$. In the latter case, it is important to also include the ac part of the tension \cite{poot_PSSB_modelling_CNT}. Although the mode shapes $\chi_n(x)$ can be solved analytically, finding the tension and eigenvalues $\omega_n$ typically has to be done numerically \cite{poot_PSSB_modelling_CNT, Hoch_Sbeam}.

\section{Projection onto modes} \label{sec:projection}
The full equations of motion [Eqs.~\eqref{eq:EulerBernoulli_u} and \eqref{eq:EulerBernoulli_v}] are partial differential equations that govern both the spatial profile and the dynamics. Getting insights from these directly is therefore not easy. The analysis can be greatly simplified by assuming a specific displacement profile and projecting onto that mode. Of course, the better that \emph{Ansatz} is, the better the agreement between the dynamics calculated using the full and the reduced equations of motion will be. 
In our experimental work \cite{Hoch_Sbeam}, the focus was on the so-called ``Sbeam'' design that is close to the cosine shape. Inspired by this, we take 
\begin{equation}
    \{u_0,u,v\}  =  \{U_0,U,V\} \times \halfl \big(1-\cos(2\pi x/L)\big) \label{eq:projection}
\end{equation}
for the pre- and post-release in-plane displacement and the out-of-plane displacement, respectively. Note that displacement profiles of the form \eqref{eq:projection} satisfy the boundary conditions for doubly-clamped beams \cite{cleland_nanomechanics} and also correspond to the shape of buckled beams \cite{poot_physrep_quantum_regime}. With a single anti-node at $x = L/2$, they also resemble the shape of the fundamental in- and out-of-plane flexural modes of beams and strings. On the other hand, Eq.~\eqref{eq:projection} is not expected to work well for higher modes which have very different shapes, e.g with more nodes. For these, different projections for the static and dynamic behaviour may be used \cite{westra_PRL_coupled}. Alternatively the full model [Eqs.~\eqref{eq:EulerBernoulli_u} and ~\eqref{eq:EulerBernoulli_v}] can be solved, or finite-element simulation can be performed. However, in the following we focus on the fundamental modes.
Inserting Eq.~\eqref{eq:projection} into the expression for $K$, $E_B$, $E_T$, and $T$, and performing the integration yields:
\begin{eqnarray}
K & = &  \half m \left[ \dot U^2 + \dot V^2 \right] \times \frac{3}{8}
\label{eq:Kprojected}
\\
E_B & = & \half \left[D_y  (U-U_0)^2 + D_z V^2 \right] \frac{1}{L^{3}} \times 2\pi^4 
\label{eq:EBprojected}
\\
E_T & = & \half \left(\halfl T_0 + \halfl T \right) \left[ U^2 + V^2 \right] \frac{1}{L} \times \halfl \pi^2 
\label{eq:ETprojected}
\\
T & = & T_0 + \frac{EA}{2L^2} \left[U^2 + V^2 \right] \times \halfl \pi^2, \label{eq:Tprojected}
\end{eqnarray}
where $m = \rho L h w$ is the total mass of the beam \cite{poot_physrep_quantum_regime}. The factors after the multiplication sign depend on the assumed displacement profile. If, for example, instead of $\halfl(1-\cos(2\pi x/L))$, $\sin(\pi x/L)$ was chosen (i.e. the fundamental mode shape of a string, as well as the pre-displacement of our ``sine'' design \cite{Hoch_Sbeam}) these factors would be $\halfl$, $\halfl \pi^4$, $\halfl \pi^2$, and $\halfl \pi^2$, respectively. In words, for the same amount of center displacement, it would have less bending energy, but equal stretching energy and tension as well as a higher effective mass $m_\mathrm{eff} = \frac{3}{8}m \rightarrow \half m$. In the following, we employ on the shape given by Eq.~\eqref{eq:projection} for the projection onto the modes; this agrees well with finite-element simulations as will be shown below.

\begin{figure}[!tbhp]
  \includegraphics[width=\columnwidth]{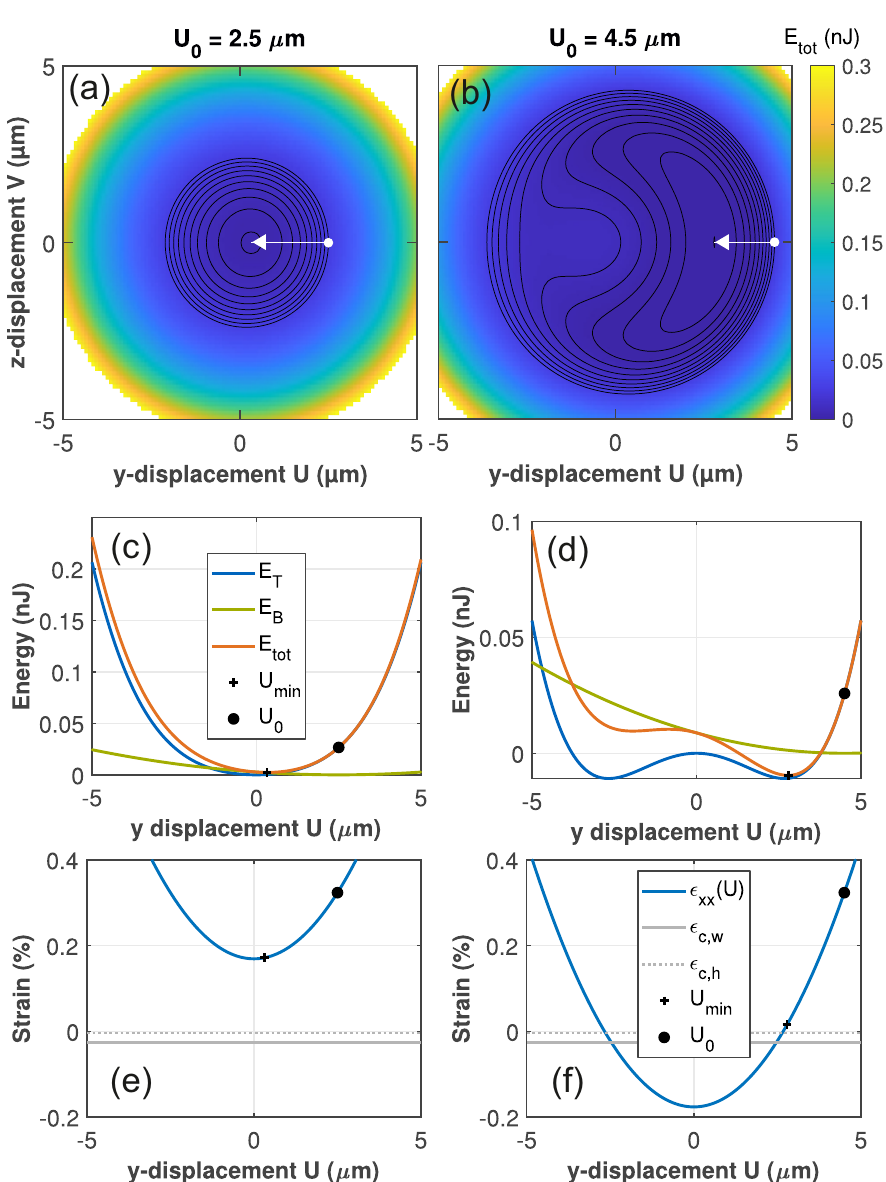}
  \caption{
  Potential energy and strain of a $L = 100 \un{\mu m}$ long beam with pre-displacement $U_0 = 2.5 \un{\mu m}$ (left) and $U_0 = 4.5 \un{\mu m}$ (right). The top panels show the energy landscape both as colormap and with contour lines. The middle (bottom) panels show the energy (strain) for $V=0$. The round symbol indicates $U_0$ and the arrow cross the minimum of $E_\mathrm{tot}$, $U_\mathrm{min}$. The complete list of parameter values are given in the Appendix \ref{app:parameters}.}
  \label{fig:potential}
\end{figure}

\subsection{Potential energy landscape}
\label{ssec:landscape}
After the projection onto the mode shape using Eq.~\eqref{eq:projection}, the total potential energy is a function of the two center displacements, $U$ and $V$. Figure~\ref{fig:potential}(a) and (b) show this two-dimensional energy landscape for beams with two different center pre-displacements $U_0$. First of all, note that both potentials are symmetric with respect to $V=0$. This can be understood because only $V^2$ appears in Eqs.~\eqref{eq:Kprojected}-\eqref{eq:Tprojected}. Moreover, both panels show that it costs energy to displace the beam beyond $U_0$ and that the potential energy can be lowered by straightening ($0\le U < U_0$). Still, they show very different behaviour: For $U_0 =  2.5 \un{\mu m}$, the potential energy contours appear almost concentric, and the relaxed displacement, given by the position of the potential minimum $U = U_\mathrm{min}$, is close to 0. Figure.~\ref{fig:potential}(c) indicates that in this case the tension energy (blue) dominates over the bending energy (green) and that after straightening the string still has a considerable positive strain value [Fig.~\ref{fig:potential}(e)]. For $U_0 =  2.5 \un{\mu m}$, the beam thus almost completely straightens $U_\mathrm{min} \ll U_0$ and still has considerable tensile stress after relaxing.

For $U_0 = 4.5 \un{\mu m}$, the situation is different. In this case, the potential landscape is more complex as both the contour plot in (b) and the line cuts in (d) show. From the latter, $E_\mathrm{tot}$ appears to have two minima, but, as panel (b) shows, the left one is actually a saddle point. The global minimum is at $U_\mathrm{min} \approx 2.79 \un{\mu m} \gg 0$, which indicates that there is still a significant bending. In other words, the beam has not fully straightened. Looking at the strain in Fig.~\ref{fig:potential}(f) clarifies that $U = 0$ would be a situation with a compressive strain that exceeds the critical value for buckling (see Sec.~\ref{ssec:buckling}). Indeed, looking back at Fig.~\ref{fig:relaxation}(c) shows that the beam with $U_0 = 2.5 \un{\mu m}$ does not have the potential to buckle, whereas the one with $U_0 = 4.5 \un{\mu m}$ has. In the latter case, the beam thus remains displaced, but has only very small (tensile) stress after relaxing. Note that there is no out-of-plane displacement, which would be $V_\mathrm{min} \neq 0$; in our numerical studies that situation was only encountered for compressive initial stress $\sigma_0 < 0$. 
Even for $D_y \gg D_z$, with tensile film stress the pre-displaced beams thus ``prefer'' an in-plane displacement above out-of plane buckling. Still, the potential to buckle as introduced in Sec. \ref{ssec:buckling} is an important parameter that indicates if the beams will almost completely straighten, or not.

\begin{figure}[!tbhp]
  \includegraphics[width=0.8\columnwidth]{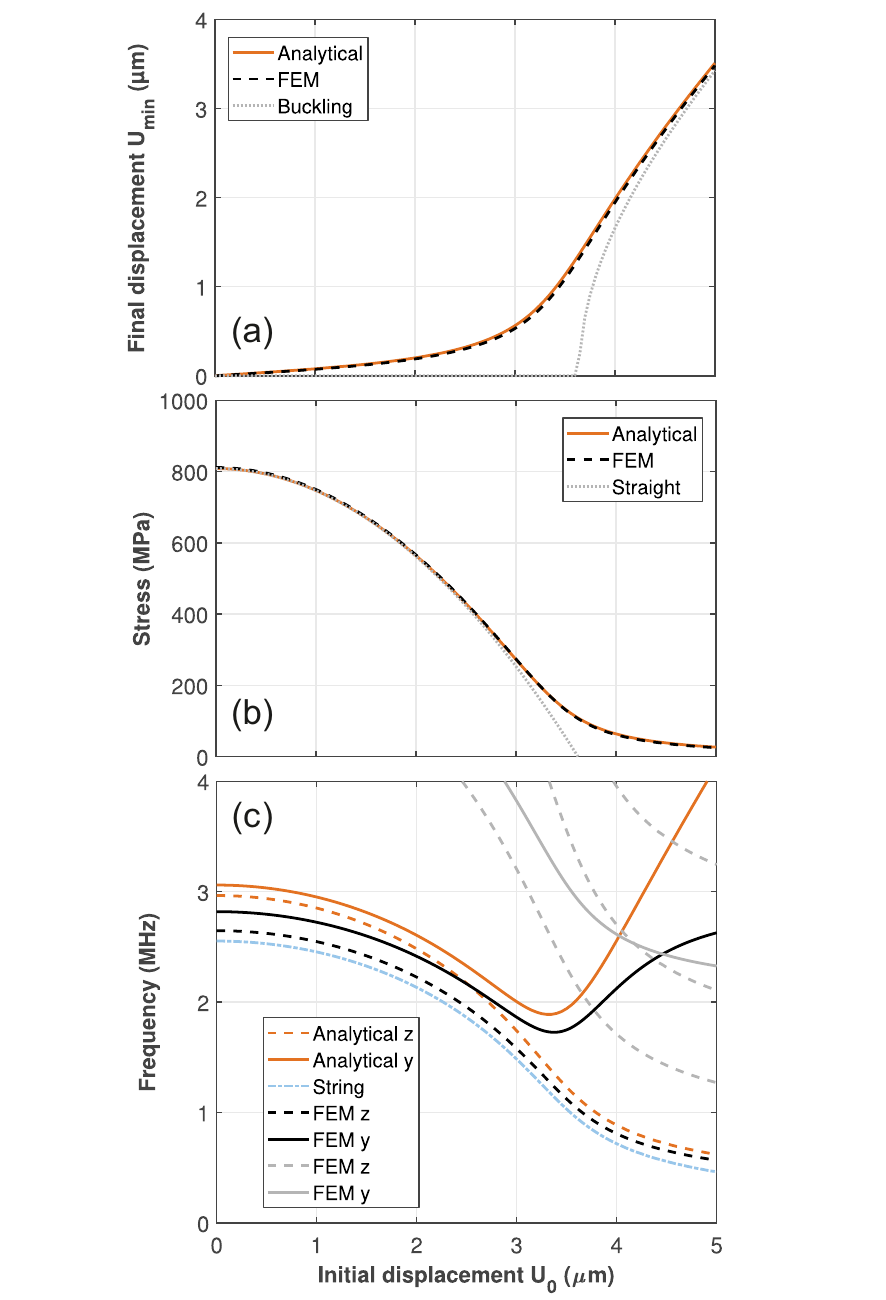}
  \caption{
  The final displacement $U_\mathrm{min}$ (a), stress (b), and resonance frequencies (c) as obtained using the analytical modal projection model (orange) and comparison with finite-element simulations (black) as a function of the pre-displacement $U_0$ of a $L=100 \un{\mu m}$ beam; 
  other parameter values are given in the Appendix \ref{app:parameters}. 
  In panel (a) the stress when the beam would be straight $T_0/A$ is also shown (dotted gray). In (b) the dotted gray line shows the displacement $2L/\pi([A\sigma_{c,w}-T_0])^{1/2}$ of a buckled beam with tension $T_0 < A\sigma_{c,w} < 0$ \cite{poot_physrep_quantum_regime}.
  In panel (c) the frequencies are compared to that of a string (light blue). There,
  in- (y) and out-of-plane (z) polarized modes are indicated with solid and dashed lines, respectively and the fundamental (higher) modes from the FEM simulations are indicated in black (gray).}
  \label{fig:vs_U0}
\end{figure}

\subsection{Pre-displacement dependence}
\label{ssec:U0}
The position of the minimum in the potential energy $U_\mathrm{min}$ can be tracked as a function of the pre-displacement $U_0$. With its value, also the relaxed tension can be calculated from Eq.~\eqref{eq:Tprojected}. Figure \ref{fig:vs_U0}(a) and (b) show $U_\mathrm{min}$ and the tension normalized by the cross sectional area, respectively. These show that, in agreement with the experiments \cite{Hoch_Sbeam}, for $U_0 \lesssim 3 \un{\mu m}$, the final displacement is close to zero and that the stress is still relatively large. The latter is close to that of a straight string $T_0/A$, as calculated using Eq.~\eqref{eq:T0} (dotted). Beyond where $T_0/A$ becomes zero, the stress becomes small - but is still tensile - and the displacement $U_\mathrm{min}$ grows. Interestingly, the relaxed displacement approaches that of a buckled beam under compressive tension $T_0/A < 0$ (dotted line) \cite{nayfeh_buckled}. This shows once more that the behaviour of the beams is intimately related to their potential to buckle. Both before, after, and close to the buckling transition, both the final stress and displacement obtained from the analytical modal-projection model are almost indistinguishable from those obtained using finite-element simulations (FEM, dashed lines). This shows that the analytical model can be used to accurately describe the static relaxation of the pre-displaced beams.

\subsection{Reduced equations of motion}
\label{ssec:reduced}
To derive the reduced equations of motion, i.e. the differential equations that govern the dynamics of the center displacements $U$ and $V$, the formalism of Hamiltonian mechanics is employed. There, Hamilton's equations relates the generalized momenta $P_U$ and $P_V$ that are associated with $U$ and $V$, respectively, to derivatives of the Hamiltonian $H = K + E_\mathrm{tot}$ with respect to said quantities, and vice versa \cite{Wells_Lagrangian_dynamics}:
\begin{eqnarray}
P_U & = & +\pder{H}{\dot U} = m_\mathrm{eff} \dot U,~
P_V   =   +\pder{H}{\dot V} = m_\mathrm{eff} \dot V
\label{eq:HamiltonP}
\\
\dot P_U & = & m_\mathrm{eff} \ddot U = -\pder{H}{U},~
\dot P_V   =   m_\mathrm{eff} \ddot V = -\pder{H}{V}.
\end{eqnarray}
From this, the reduced equations of motion follow directly:
\begin{eqnarray}
m_\mathrm{eff} \ddot U & = & -2\pi^4 D_y(U-U_0)/L^3 - \halfl \pi^2 T[U,V] U /L \nonumber \\
 & & \hspace{0.5cm} - EA \pi^4 U^3/8L^3 
\label{eq:HamiltonU}
\\
m_\mathrm{eff} \ddot V & = & -2\pi^4 D_z V/L^3 - \halfl \pi^2 T[U,V] V /L \nonumber \\
 & & \hspace{0.5cm} - EA \pi^4 V^3/8L^3.
\label{eq:HamiltonV}
\end{eqnarray}
Here, $m_\mathrm{eff} = \frac{3}{8} m$ is again the effective mass. Comparing these equations to the equation of motion for a harmonic oscillator shows that the part of the right hand sides that is proportional to $U$ and $V$, respectively, contains the spring constants $k_{y,z}$ and these determine the resonance frequencies. Both the first (i.e. bending) and second (tension, via $T_0$) term contribute to this. Besides these linear contributions, there are e.g. also terms proportional to $U^3$ and $UV^2$. 
The last term in Eqs.~\eqref{eq:HamiltonU} and \eqref{eq:HamiltonV} is clearly nonlinear, but, since the tension $T$ depends on $U$ and $V$ [see Eq.~\eqref{eq:Tprojected}], also the second term contributes to the beam's nonlinearities. This term also nonlinearly couples in- and out-of-plane motion  \cite{westra_PRL_coupled, nayfeh_nonlinear}; a detailed analysis of these nonlinear effects in pre-displaced beams will, however, be published elsewhere. 

By linearizing Eqs.~\eqref{eq:HamiltonU} and \eqref{eq:HamiltonV} around $(U_\mathrm{min}, V_\mathrm{min} = 0)$, the spring constants $k_{y,z}$ are obtained and from these the eigenfrequencies $f = \sqrt{k_{y,z}/m_\mathrm{eff}}/2\pi$ \footnote{Note that due to the symmetry with respect to $V=0$ (see Sec.~\ref{ssec:landscape}), there is no linear coupling ($\pderl{^2E_\mathrm{tot}}{U\partial V} = 0$ for $V=0$) between $U$ and $V$ and the eigenmodes are purely y and z polarized.}.
Figure \ref{fig:vs_U0}(c) shows the frequencies calculated using the analytical modal projection model as a function of $U_0$ in orange. Both the frequency for the z and for the y polarized modes show good agreement with those calculated using finite-element simulations. The vertical offset between the analytical model and the finite-element simulations can be explained by the difference between the assumed and the actual mode shape. For small $U_0$, both frequencies also follow the same trend as the frequency of a string (light blue) \cite{Verbridge_JournalofAppliedPhysics_highstress_SiN, Schmid_PhysRevB_damping_model, Ghadimi_NanoLett_SiN_loss_engineering} when using the tensile stress calculated with Eq.~\eqref{eq:Tprojected} [cf. the orange line Fig.~\ref{fig:vs_U0}(a)]. For larger $U_0$, the z-polarized mode continues to follow the frequency of a string, but the y-polarized mode has a different behaviour, both in the analytical model and the finite-element simulations. 
As shown in Sec.~\ref{ssec:shapes}, especially during the upward trend of the in-plane mode with $U_0$, that in-plane mode shape is strongly modified, explaining why the deviation between the reduced model and the FEM simulation increases there. Still, the upward trend and the position of the transition are reproduced by the analytical model. The modal projection can thus be also used to understand the dynamics of the fundamental modes. For example, the difference between the in and out-of-plane mode frequencies for large $U_0$ can be directly related to the ellipsoidal equipotential contours in Fig.~\ref{fig:potential}(b) around the minimum, which indicate that the curvature of $U_\mathrm{tot}$, i.e. the spring constants $k_{y,z}$, are very different for the $V$ and $U$ direction.

\section{Finite-element simulations} \label{sec:FEM}
To go beyond the analytical model presented in the previous sections and the Euler-Bernouli equations of Eqs.~\eqref{eq:EulerBernoulli_u} and \eqref{eq:EulerBernoulli_v}, we also performed finite-elements (FEM) simulations using COMSOL MultiPhysics\textsuperscript{\textregistered}. The model is built using the Solid Mechanics toolbox, and the control of geometric parameters, as well as the extraction of the results, is done with \textsc{Matlab}\textsuperscript{\textregistered} via the LiveLink\texttrademark~interface as detailed in \cite{Hoch_Sbeam}. From the FEM simulations, both static and dynamic quantities can be obtained, as was shown in Fig. \ref{fig:vs_U0}. There, there was a very good agreement between the projected model and the FEM simulations for both the final displacement and the stress. For the eigenfrequencies in Fig. \ref{fig:vs_U0}(c), deviations between the analytical model and the simulations were visible, that were attributed to difference between the actual mode shape and the assumed cosine-shape of Eq. \eqref{eq:projection}. In the following the exact mode shape will be studied in more detail using FEM simulations. Also the role of the overhanging clamping points and the stress distribution within the beams will be studied in this Section.

\subsection{Mode shapes}\label{ssec:shapes}
\begin{figure}[!tbhp]
  \includegraphics[width=1\columnwidth]{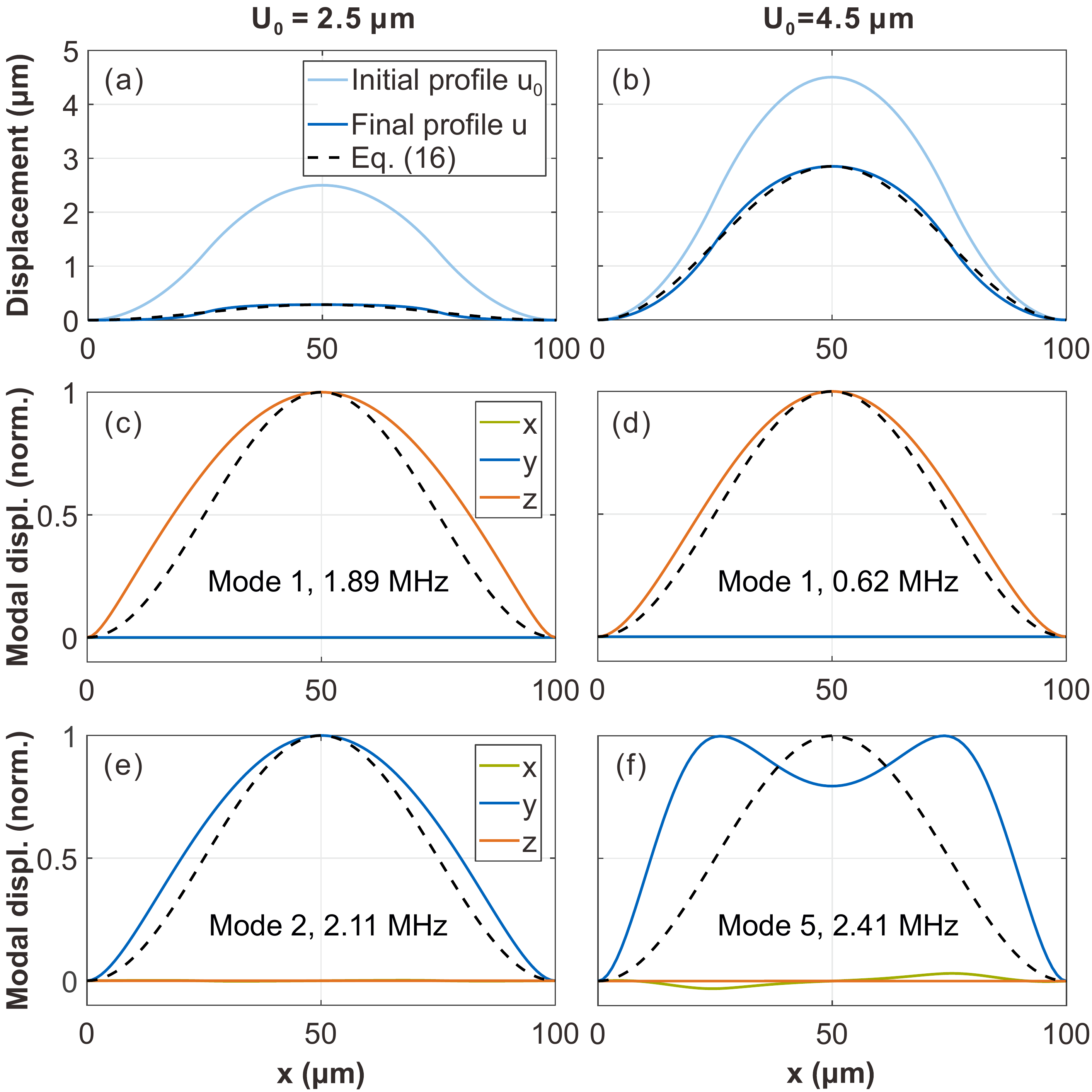}
  \caption{
  Static profiles and eigenmodes of $L=100 \un{\mu m}$ beams obtained with FEM simulations. Left panels are for $U_0=2.5 \un{\mu m}$ and the right for $4.5 \un{\mu m}$. (a) and (b) show the static displacement profile before ($u_0(x)$, light blue) and after release ($u(x)$, blue). Corresponding normalized mode shapes of the fundamental z-polarized (c, d) and y-polarized (e, f) eigenmodes are illustrated, respectively. In all panels, the dashed black line indicates the cosine shape from Eq.~\eqref{eq:projection} with the same maximum displacement. The different colors correspond to the different Cartesian components of the displacement vector.}
  \label{fig:Mode_simu}
\end{figure}

To validate the \emph{Ansatz} made in Sec.~\ref{sec:projection}, first the static beam shape before and after relaxation is investigated. For this, the geometry of the Sbeam is created as detailed in Ref.~\cite{Hoch_Sbeam} and its static relaxation is computed. 
As shown in Fig.~\ref{fig:Mode_simu}(a) and (b), the two typical pre-displaced beams that were also studied analytically, both straighten after relaxing, i.e. their final profiles $u(x)$ (blue) are smaller than the initial profile $u_0(x)$. The Sbeam is nearly straight for $U_\mathrm{0}=2.5 \un{\mu m}$ while still having a significant displacement remaining for $4.5 \un{\mu m}$, which confirms the discussion from Fig.~\ref{fig:vs_U0}(a). The relaxed static profile is also compared with the cosine function of Eq.~\eqref{eq:projection} (dashed black line). Indeed, the simulated $u(x)$ is described well by that \emph{Ansatz}. 

Next, the eigenmodes are simulated and Fig.~\ref{fig:Mode_simu}(c)-(f) shows the first out-of-plane (``Z1'') and in-plane (``Y1'') polarized modes for $U_0=2.5 \un{\mu m}$ (left) and $4.5 \un{\mu m}$ (right). The simulated eigenfrequencies of these modes are also indicated.
In both cases, the lowest eigenmode is the fundamental out-of-plane mode Z1. Taking a closer look at the mode profile shows clear differences between the two pre-displacements: The left one has finite slope close to the clamping points and looks thus more like a sin shape of a pure string rather than a cosine. That is due to the significant remaining stress after relaxing (see Fig.~\ref{fig:vs_U0}(b)) \cite{poot_physrep_quantum_regime}. By comparison, the lower final tension for $U_\mathrm{0}=4.5 \un{\mu m}$ gives a larger bending contribution resulting in a much more rounded shape, which is captured well by the cosine function.

For $U_\mathrm{0}=2.5 \un{\mu m}$, the in-plane mode shape (e) looks similar to that of the out-of-plane mode and also matches well with the cosine shape (dashed line) of Eq.~\eqref{eq:projection}. However, the Y1 mode of the $U_\mathrm{0}=4.5 \un{\mu m}$ beam shape is different (f). The central maximum in the modal displacement is now a local minimum. Instead, two maxima appear near one and three quarter of $L$. Such mode shapes are characteristic of buckled beams (see e.g. \cite{nayfeh_nonlinear} and \cite{etaki_natphys_squid}). Note, that Fig. \ref{fig:vs_U0}(c) showed that at $4.5 \un{\mu m}$ pre-displacement, the mode of Fig.~\ref{fig:Mode_simu}(f) has crossed the y-polarized mode with a single node (``Y2''), similar what happens for the aforementioned buckled beams. Hence at point the Y1 mode is actually the fifth eigenmode of the structure and lies above the odd Y2 mode. Still, we stick to this nomenclature as the mode in Fig.~\ref{fig:Mode_simu}(f) is a direct continuation of the original Y1 at $U_0 = 0$.

All these considerations indicate that a small tension, or equivalently a large remaining $U$, can impact the in-plane eigenmodes due the close connection to the dynamics of buckled beams. Note that in this case, the mode clearly deviates from the assumed cosine shape so that in this regime the analytical model from Sec. \ref{sec:projection} is no longer accurate, explaining the deviations in Fig. \ref{fig:vs_U0}(c) between that model and the FEM simulations. Still for most of the parameters, the simulated fundamental mode shape is describes  to a good approximation by Eq.~\eqref{eq:projection}.

\subsection{Role of overhang}\label{ssec:overhang}
Another important question is how the details of the clamping region influence the statics and dynamics of the pre-displaced beams. In the experiments \cite{Hoch_Sbeam}, the beams are defined by vertically etching the structures into the silicon nitride \cite{Terrasanta_OE_AlN_on_SiN}, followed by  isotropic etching of the silicon oxide underneath. This causes an ``overhang'' of the clamping region, which can have an effect on e.g. the residual stress and resonance frequencies \cite{buckle_PRAppl_tension_universal_length_dependence, babei_gavan_JMM_undercut}. The size of this overhang $O$ is determined by the depth of the isotropic silicon-oxide etch and is $\sim 660 \un{nm}$ in our experimental realization \cite{Hoch_Sbeam}.
\begin{figure}[!tbhp]
  \includegraphics[width=1\columnwidth]{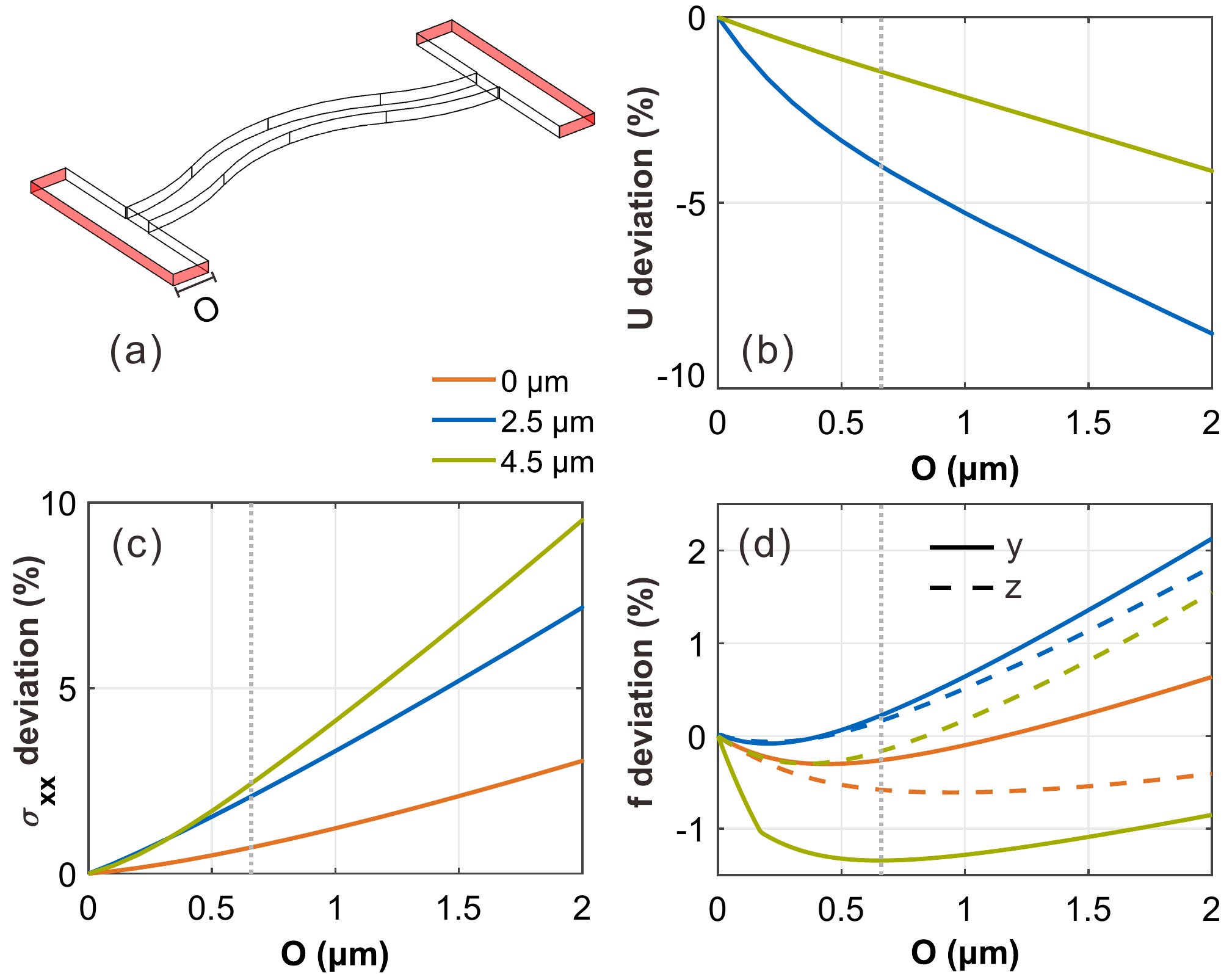}
  \caption{Effects of overhang on three different pre-displaced Sbeams' stationary and dynamical behavior. All have $L=100 \un{\mu m}$. 
  (a) Schematic of the geometry of an Sbeam with two extra overhanging clamping regions. The fixed boundaries are indicated by red color. 
  (b) The relative change of the center displacement to the case without overhang (i.e., $[U(O)-U(O=0)]/U(O=0)\times100\%$). 
  (c) Relative change of the stress component $\sigma_{xx}$. 
  (d) Relative change of the first in- (solid lines) and out-of-plane (dashed lines) eigenfrequencies. 
  The colors correspond to different pre-displacements $U_0$ and are consistent between the panels. The dotted gray lines indicate the etch depth of $\sim 660 \un{nm}$ in our experiments. 
\cite{Hoch_Sbeam}.
  \label{fig:Overhang}}
\end{figure}

To model the role of such an overhang, the FEM geometry of the pre-displaced beams is extended with two rectangular pads that are clamped at their outside sides, as shown in Fig.~\ref{fig:Overhang}(a). Their extent in the y-direction is chosen large enough that the exact boundary condition at those ends does not influence the results. Intuitively, the overhang will make the clamping of the beam less rigid compared to the case with fixed boundary conditions at $x=0,L$. This will affect the final displacement, stress, and resonance frequencies. In accordance with the previous sections, here the center displacement $U$ is defined as the y-component of the displacement vector field $\vec{u}(x,y,z)$ evaluated at the beam center: $U(O) = u_\mathrm{y}(\frac{L}{2}, U_\mathrm{0}, 0)|_{\mathrm{Overhang}=O}$ \footnote{The Cartesian coordinates used here correspond to the original, undeformed geometry.}. Figure~\ref{fig:Overhang}(b) shows that for both pre-displacements $2.5$ and $4.5 \un{\mu m}$, the relative change in $U$ is negative, indicating that the relaxed beams retain less final displacement when the overhang gets bigger (and for $U_0=0$, it stays $0$). Furthermore, the residual stress component $\sigma_{xx}$ shown in Fig.~\ref{fig:Overhang}(c) increases with the overhang. The larger $U_{0}$, the higher this relative increase of $\sigma_\mathrm{xx}$ is, which is consistent with the model by B\"uckle \emph{et al.}  \cite{buckle_PRAppl_tension_universal_length_dependence}. 
In this case, the tension in the wide overhang regions pulls on the beam, thereby increasing the tension of the latter \cite{Bereyhi_NanoLett_clamp_tapering}.

The eigenfrequencies also have a relative shift compared to a beam without overhang as Fig.~\ref{fig:Overhang}(d) shows. However, the $f$ change is nonmonotonic as it exhibits both downward and upward shifts. That can be understood from the interplay between the increase in effective length \cite{babei_gavan_JMM_undercut} on the one hand and the increase in the stress on the other hand. The former leads to a reduction in the frequency of flexural modes, whereas the latter increases the frequency \cite{buckle_PRAppl_tension_universal_length_dependence, Bereyhi_NanoLett_clamp_tapering}. The competition between these two effect leads to the complex behaviour observed in Fig.~\ref{fig:Overhang}(d), where, depending on displacement polarization, pre-displacement, and overhang, both a positive or a negative frequency shift can be obtained.
Still, the experimental etch depth of $\sim 660 \un{nm}$  \cite{Hoch_Sbeam} will only cause small shifts of frequencies of less than $\sim 1\%$ for $100 \un{\mu m}$ string. For the stress and final displacement, the changes are slightly larger but never exceed the percent level so that in many cases its effect can be neglected. Nevertheless, for very accurate modelling of the strings, the overhang should be included.

\subsection{Bending and stress distribution}
The beam geometry has $L \gg w, h$ and in the above discussions, the beam was viewed as a one- (Sec.~\ref{sec:EoM}) or even zero-dimensional object (Sec.~\ref{sec:projection}) without considering its cross sections. Still, it is important to look how the stress of each continuum element \cite{nayfeh_nonlinear} is re-distributed along the beam and over its cross section after relaxing, especially to understand better why the Sbeam does not become fully straight. Here, only the xx-component is shown because the FEM simulations indicate that all other components are orders of magnitude of smaller ($< 1 \un{MPa}$) and thus negligible compared to $\sigma_{xx}$, even at the largest geometric stress tuning. This also confirms the argument in Sec.~\ref{sec:relaxation}.
\begin{figure}[!tbhp]
  \includegraphics[width=1\columnwidth]{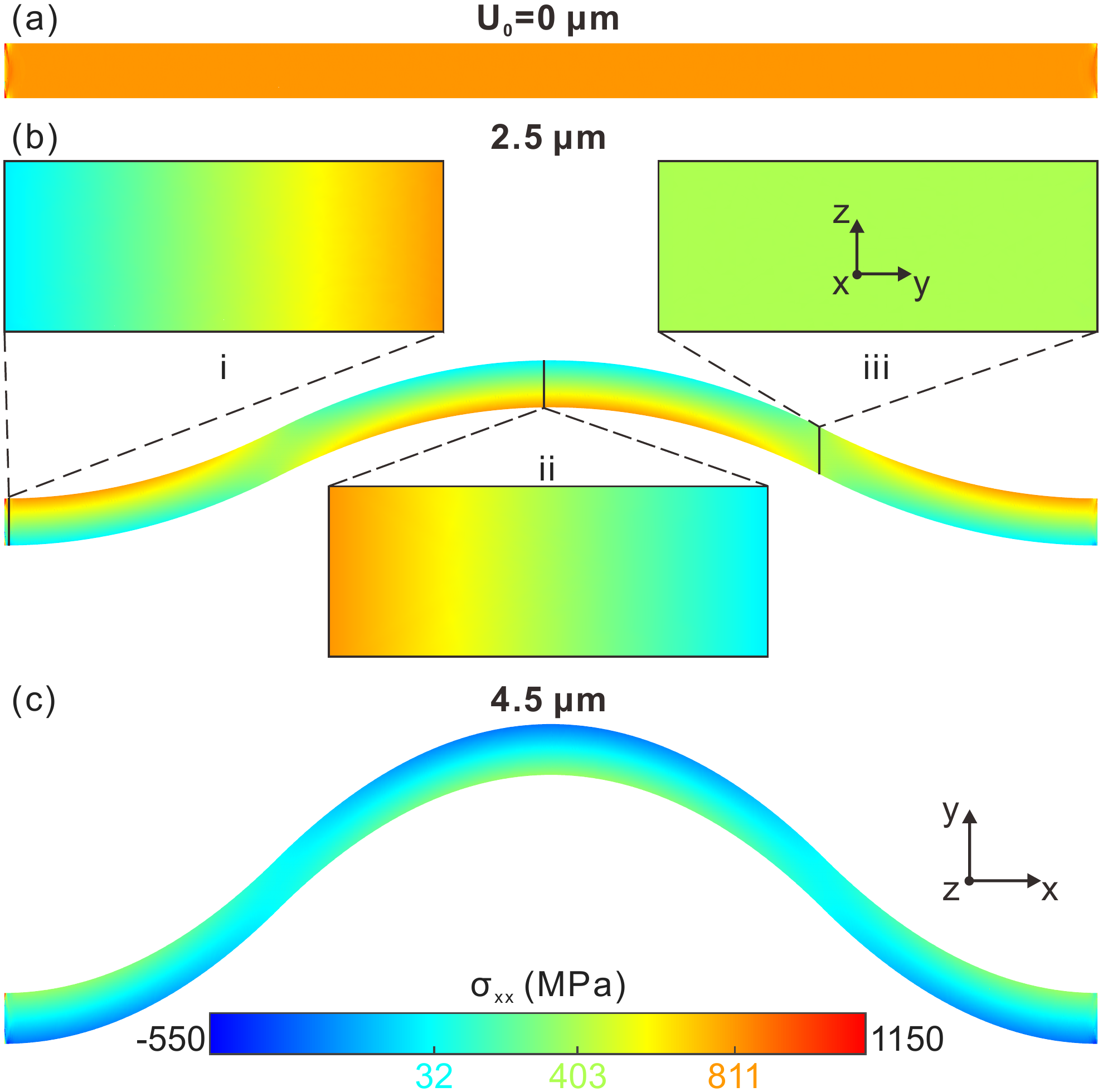}
  \caption{Distribution of the stress component $\sigma_{xx}$  for (a) $U_{0}=0$, (b) $2.5$, and (c) $4.5 \un{\mu m}$ along the beam (xy-plane). The average $\sigma_{xx}$ over beam is $811$, $403$, and $32$ \un{MPa}, respectively, as indicated in the colorbar. For panel (b) also the stress distribution at yz cut planes are shown near a clamping point (i), halfway (ii), and at a quarter $L$ (iii). For clarity, xy-cuts have been plotted with a different vertical and horizontal scale. This, may give the impression that the difference between $\ell[u_0,0]$ and $L$ and the angle between the center curve and the x axis are large, but in reality these are only $(\ell-L)/L = 0.17~(0.54) \% $ and $0.10~(0.18)\un{rad}$ for $U_0 = 2.5~(4.5) \un{\mu m}$, respectively.
  \label{fig:stress_distribution}}
\end{figure}

Figure~\ref{fig:stress_distribution} shows the $\sigma_\mathrm{xx}$ distribution of three different pre-displaced beams; the values averaged over the entire beam are indicated in the colorbar. In agreement with Fig.~\ref{fig:vs_U0}(b), the larger $U_\mathrm{0}$, the smaller the average final stress. However, these average values does not tell the entire story. The xy projections show that locally the stress can deviate from the mean. Although for the straight beam [panel (a)] the average of $\sigma_\mathrm{xx}$ is $811 \un{MPa}$ is distributed uniformly over the structure, the stress is nonuniform for the pre-displaced beams [(b) and (c)], even reaching significant negative (i.e. compressive) values for $U_\mathrm{0}=4.5 \un{\mu m}$. First, note that in all cases $\sigma_{xx}$ is symmetric about the center $x=\frac{L}{2}$ Moreover, the stress distributions at three yz-planes are plotted for $U_\mathrm{0}=2.5 \un{\mu m}$ [(i), (ii), and (iii)], showing that, as expected for thin beams \cite{cleland_nanomechanics, poot_physrep_quantum_regime}, the stress is constant along the thickness of the beam. Therefore, cuts at constant $z$ are sufficient for fully representing the stress distribution in these structures.
%

The xy-cuts in Fig.~\ref{fig:stress_distribution}(b) and (c) show that for the two pre-displaced beams $\sigma_{xx}$ is different on both sides of the center curve (which corresponds to the so-called neutral plane \cite{cleland_nanomechanics}). 
Overall, the relaxation is accompanied by a shrinking of the beam's curve length as discussed in Sec.~\ref{sec:relaxation}, causing the reduction in the average longitudinal stress. On top of this global effect, one side of the beam is stretched more than average, and thus has a higher-than-average stress, whereas the other side is stretched less or even compressed, resulting in a lower local $\sigma_{xx}$ that can even be compressive ($\sigma_{xx} < 0$, blue). 
Figure~\ref{fig:stress_distribution} shows that from 0 to $L/4$ and from $3L/4$ to $L$, the stress is higher at the upper edge of the xy-cut and lower at the lower edge. This situation is reversed between $L/4$ and $3L/4$. The difference between these two regions coincides with the inflection point of the center curve. The sign of the local curvature of $u_0$ thus determines whether the upper or lower side has a higher-than-average stress. A more detailed analysis of the distribution of $\sigma_{xx}$ over the beam width indicates a linear dependence on $y$ around the average value. This is exactly as expected for bending of the beam \cite{cleland_nanomechanics, poot_physrep_quantum_regime}. The FEM simulations thus confirm that the final relaxation is determined by the interplay between the tension and bending rigidity as predicted by our analytical model. Finally, it should be kept in mind that for a static situation, the longitudinal stress integrated over the width and thickness of the beam is constant along $x$. This can be understood since any variation of the tension $T \equiv \iint \sigma_{xx} \intd y \intd z$ leads to longitudinal displacements that will balance the gradient in $T$ \cite{flensberg_NJP_electronphonon}.



\section{Conclusion}
A theoretical framework to analyze the relaxation and dynamics of pre-displaced beams was presented. First, the relaxation of straight and pre-displaced beams was studied and expressions for the bending and tension energy were derived. For the tension energy, it is neither the initial nor the final tension that appears, but their average. The equations of motion were derived and a modified Euler-Bernoulli equation is obtained. The pre-displacement appears as an additional in-plane force. In the limit of high tension, the resonators behave as simple strings with a geometrically-tunable tension. By projecting on the fundamental mode shape, the system is reduced to two variables: the in- and out-of-plane displacements at the center. From the energy landscape, insights in the relaxation and the role of buckling are obtained. This reduced model can be used to understand the static relaxation and dynamics of the fundamental modes, such as the geometric tuning of the stress and resonance frequency. Finally, the analytical model is supported by finite-element simulations of the mode shapes, the role of the overhang, and stress profiles. This enables a good understanding of the experimental observations in Ref. \cite{Hoch_Sbeam} and future work will explore the nonlinear properties of the pre-displaced beams more detail.

\appendix*
\section{Parameter values}
\label{app:parameters}
The parameters used are inspired by our experimental work described in Ref.~\cite{Hoch_Sbeam}, where high-stress silicon nitride beams were used. The nominal values of the parameters used in the calculations and simulation in this Article are given in Table \ref{tab:parameters}.

\begin{table}[!h]
  \centering
  \caption{Parameter values used for the calculations (unless stated otherwise).}
    \begin{tabular}{cccc}
   \hline
    Parameter & Description & Value & Source \\
    \hline
    $h$  & Thickness & $330 \un{nm}$ & \cite{Hoch_Sbeam}\\
    $w$ & Width & $850 \un{nm}$ & \cite{Hoch_Sbeam}\\
    $L$ & Length & $100 \un{\mu m}$ & \cite{Hoch_Sbeam} \\
    $\sigma_\mathrm{film}$ & Film stress & $ 1050.1 \un{MPa}$ & \cite{hoch_MM_mode_mapping} \\
    $\rho$ & Density & $3.10 \times 10^3\un{kg/m^3}$ & \cite{COMSOL_SiN} \\
    $E$ & Young's modulus & $250 \un{GPa}$ & \cite{COMSOL_SiN} \\
    $\nu$ & Poisson ratio & $0.23$ & \cite{COMSOL_SiN} \\
    \hline
    \end{tabular}%
  \label{tab:parameters}%
\end{table}%

\begin{acknowledgments}
This research was funded by the German Research Foundation (DFG) under Germany's Excellence Strategy - EXC-2111-390814868 and TUM-IAS, which is funded by the German Excellence Initiative and the European Union Seventh Framework Programme under grant agreement 291763. We thank Timo Sommer and Pedro Soubelet for discussion.
\end{acknowledgments}

\bibliography{Sbeam}

\begin{thebibliography}{46}%
\makeatletter
\providecommand \@ifxundefined [1]{%
 \@ifx{#1\undefined}
}%
\providecommand \@ifnum [1]{%
 \ifnum #1\expandafter \@firstoftwo
 \else \expandafter \@secondoftwo
 \fi
}%
\providecommand \@ifx [1]{%
 \ifx #1\expandafter \@firstoftwo
 \else \expandafter \@secondoftwo
 \fi
}%
\providecommand \natexlab [1]{#1}%
\providecommand \enquote  [1]{``#1''}%
\providecommand \bibnamefont  [1]{#1}%
\providecommand \bibfnamefont [1]{#1}%
\providecommand \citenamefont [1]{#1}%
\providecommand \href@noop [0]{\@secondoftwo}%
\providecommand \href [0]{\begingroup \@sanitize@url \@href}%
\providecommand \@href[1]{\@@startlink{#1}\@@href}%
\providecommand \@@href[1]{\endgroup#1\@@endlink}%
\providecommand \@sanitize@url [0]{\catcode `\\12\catcode `\$12\catcode
  `\&12\catcode `\#12\catcode `\^12\catcode `\_12\catcode `\%12\relax}%
\providecommand \@@startlink[1]{}%
\providecommand \@@endlink[0]{}%
\providecommand \url  [0]{\begingroup\@sanitize@url \@url }%
\providecommand \@url [1]{\endgroup\@href {#1}{\urlprefix }}%
\providecommand \urlprefix  [0]{URL }%
\providecommand \Eprint [0]{\href }%
\providecommand \doibase [0]{https://doi.org/}%
\providecommand \selectlanguage [0]{\@gobble}%
\providecommand \bibinfo  [0]{\@secondoftwo}%
\providecommand \bibfield  [0]{\@secondoftwo}%
\providecommand \translation [1]{[#1]}%
\providecommand \BibitemOpen [0]{}%
\providecommand \bibitemStop [0]{}%
\providecommand \bibitemNoStop [0]{.\EOS\space}%
\providecommand \EOS [0]{\spacefactor3000\relax}%
\providecommand \BibitemShut  [1]{\csname bibitem#1\endcsname}%
\let\auto@bib@innerbib\@empty
\bibitem [{\citenamefont {Westerveld}\ \emph {et~al.}(2021)\citenamefont
  {Westerveld}, \citenamefont {Mahmud-Ul-Hasan}, \citenamefont {Shnaiderman},
  \citenamefont {Ntziachristos}, \citenamefont {Rottenberg}, \citenamefont
  {Severi},\ and\ \citenamefont
  {Rochus}}]{westerveld_natphot_ultrasound_sensor}%
  \BibitemOpen
  \bibfield  {author} {\bibinfo {author} {\bibfnamefont {W.~J.}\ \bibnamefont
  {Westerveld}}, \bibinfo {author} {\bibfnamefont {M.}~\bibnamefont
  {Mahmud-Ul-Hasan}}, \bibinfo {author} {\bibfnamefont {R.}~\bibnamefont
  {Shnaiderman}}, \bibinfo {author} {\bibfnamefont {V.}~\bibnamefont
  {Ntziachristos}}, \bibinfo {author} {\bibfnamefont {X.}~\bibnamefont
  {Rottenberg}}, \bibinfo {author} {\bibfnamefont {S.}~\bibnamefont {Severi}},\
  and\ \bibinfo {author} {\bibfnamefont {V.}~\bibnamefont {Rochus}},\
  }\bibfield  {title} {\bibinfo {title} {Sensitive, small, broadband and
  scalable optomechanical ultrasound sensor in silicon photonics},\ }\href
  {https://doi.org/10.1038/s41566-021-00776-0} {\bibfield  {journal} {\bibinfo
  {journal} {Nat. Photonics}\ }\textbf {\bibinfo {volume} {15}},\ \bibinfo
  {pages} {341} (\bibinfo {year} {2021})}\BibitemShut {NoStop}%
\bibitem [{\citenamefont {Calleja}\ \emph {et~al.}(2012)\citenamefont
  {Calleja}, \citenamefont {Kosaka}, \citenamefont {Paulo},\ and\ \citenamefont
  {Tamayo}}]{calleja_rscnanoscale_nanomech_sensing}%
  \BibitemOpen
  \bibfield  {author} {\bibinfo {author} {\bibfnamefont {M.}~\bibnamefont
  {Calleja}}, \bibinfo {author} {\bibfnamefont {P.~M.}\ \bibnamefont {Kosaka}},
  \bibinfo {author} {\bibfnamefont {{\'{A}}.~S.}\ \bibnamefont {Paulo}},\ and\
  \bibinfo {author} {\bibfnamefont {J.}~\bibnamefont {Tamayo}},\ }\bibfield
  {title} {\bibinfo {title} {Challenges for nanomechanical sensors in
  biological detection},\ }\href {https://doi.org/10.1039/c2nr31102j}
  {\bibfield  {journal} {\bibinfo  {journal} {Nanoscale}\ }\textbf {\bibinfo
  {volume} {4}},\ \bibinfo {pages} {4925} (\bibinfo {year} {2012})}\BibitemShut
  {NoStop}%
\bibitem [{\citenamefont {Waggoner}\ and\ \citenamefont
  {Craighead}(2007)}]{Waggoner_rsc_micronanomech_sensing}%
  \BibitemOpen
  \bibfield  {author} {\bibinfo {author} {\bibfnamefont {P.~S.}\ \bibnamefont
  {Waggoner}}\ and\ \bibinfo {author} {\bibfnamefont {H.~G.}\ \bibnamefont
  {Craighead}},\ }\bibfield  {title} {\bibinfo {title} {Micro- and
  nanomechanical sensors for environmental, chemical, and biological
  detection},\ }\href {https://doi.org/10.1039/b707401h} {\bibfield  {journal}
  {\bibinfo  {journal} {Lab Chip}\ }\textbf {\bibinfo {volume} {7}},\ \bibinfo
  {pages} {1238} (\bibinfo {year} {2007})}\BibitemShut {NoStop}%
\bibitem [{\citenamefont {Lauk}\ \emph {et~al.}(2020)\citenamefont {Lauk},
  \citenamefont {Sinclair}, \citenamefont {Barzanjeh}, \citenamefont {Covey},
  \citenamefont {Saffman}, \citenamefont {Spiropulu},\ and\ \citenamefont
  {Simon}}]{lauk2020perspectives}%
  \BibitemOpen
  \bibfield  {author} {\bibinfo {author} {\bibfnamefont {N.}~\bibnamefont
  {Lauk}}, \bibinfo {author} {\bibfnamefont {N.}~\bibnamefont {Sinclair}},
  \bibinfo {author} {\bibfnamefont {S.}~\bibnamefont {Barzanjeh}}, \bibinfo
  {author} {\bibfnamefont {J.~P.}\ \bibnamefont {Covey}}, \bibinfo {author}
  {\bibfnamefont {M.}~\bibnamefont {Saffman}}, \bibinfo {author} {\bibfnamefont
  {M.}~\bibnamefont {Spiropulu}},\ and\ \bibinfo {author} {\bibfnamefont
  {C.}~\bibnamefont {Simon}},\ }\bibfield  {title} {\bibinfo {title}
  {Perspectives on quantum transduction},\ }\href@noop {} {\bibfield  {journal}
  {\bibinfo  {journal} {Quantum Science and Technology}\ }\textbf {\bibinfo
  {volume} {5}},\ \bibinfo {pages} {020501} (\bibinfo {year}
  {2020})}\BibitemShut {NoStop}%
\bibitem [{\citenamefont {O'Connell}\ \emph {et~al.}(2010)\citenamefont
  {O'Connell}, \citenamefont {Hofheinz}, \citenamefont {Ansmann}, \citenamefont
  {Bialczak}, \citenamefont {Lenander}, \citenamefont {Lucero}, \citenamefont
  {Neeley}, \citenamefont {Sank}, \citenamefont {Wang}, \citenamefont {Weides},
  \citenamefont {Wenner}, \citenamefont {Martinis},\ and\ \citenamefont
  {Cleland}}]{oconnell_nature_quantum_piezo_resonator}%
  \BibitemOpen
  \bibfield  {author} {\bibinfo {author} {\bibfnamefont {A.~D.}\ \bibnamefont
  {O'Connell}}, \bibinfo {author} {\bibfnamefont {M.}~\bibnamefont {Hofheinz}},
  \bibinfo {author} {\bibfnamefont {M.}~\bibnamefont {Ansmann}}, \bibinfo
  {author} {\bibfnamefont {R.~C.}\ \bibnamefont {Bialczak}}, \bibinfo {author}
  {\bibfnamefont {M.}~\bibnamefont {Lenander}}, \bibinfo {author}
  {\bibfnamefont {E.}~\bibnamefont {Lucero}}, \bibinfo {author} {\bibfnamefont
  {M.}~\bibnamefont {Neeley}}, \bibinfo {author} {\bibfnamefont
  {D.}~\bibnamefont {Sank}}, \bibinfo {author} {\bibfnamefont {H.}~\bibnamefont
  {Wang}}, \bibinfo {author} {\bibfnamefont {M.}~\bibnamefont {Weides}},
  \bibinfo {author} {\bibfnamefont {J.}~\bibnamefont {Wenner}}, \bibinfo
  {author} {\bibfnamefont {J.~M.}\ \bibnamefont {Martinis}},\ and\ \bibinfo
  {author} {\bibfnamefont {A.~N.}\ \bibnamefont {Cleland}},\ }\bibfield
  {title} {\bibinfo {title} {Quantum ground state and single-phonon control of
  a mechanical resonator},\ }\href {http://dx.doi.org/10.1038/nature08967}
  {\bibfield  {journal} {\bibinfo  {journal} {Nature}\ }\textbf {\bibinfo
  {volume} {464}},\ \bibinfo {pages} {697} (\bibinfo {year}
  {2010})}\BibitemShut {NoStop}%
\bibitem [{\citenamefont {Chan}\ \emph {et~al.}(2011)\citenamefont {Chan},
  \citenamefont {Alegre}, \citenamefont {Safavi-Naeini}, \citenamefont {Hill},
  \citenamefont {Krause}, \citenamefont {Gr{\"o}blacher}, \citenamefont
  {Aspelmeyer},\ and\ \citenamefont {Painter}}]{chan2011laser}%
  \BibitemOpen
  \bibfield  {author} {\bibinfo {author} {\bibfnamefont {J.}~\bibnamefont
  {Chan}}, \bibinfo {author} {\bibfnamefont {T.}~\bibnamefont {Alegre}},
  \bibinfo {author} {\bibfnamefont {A.~H.}\ \bibnamefont {Safavi-Naeini}},
  \bibinfo {author} {\bibfnamefont {J.~T.}\ \bibnamefont {Hill}}, \bibinfo
  {author} {\bibfnamefont {A.}~\bibnamefont {Krause}}, \bibinfo {author}
  {\bibfnamefont {S.}~\bibnamefont {Gr{\"o}blacher}}, \bibinfo {author}
  {\bibfnamefont {M.}~\bibnamefont {Aspelmeyer}},\ and\ \bibinfo {author}
  {\bibfnamefont {O.}~\bibnamefont {Painter}},\ }\bibfield  {title} {\bibinfo
  {title} {Laser cooling of a nanomechanical oscillator into its quantum ground
  state},\ }\href@noop {} {\bibfield  {journal} {\bibinfo  {journal} {Nature}\
  }\textbf {\bibinfo {volume} {478}},\ \bibinfo {pages} {89} (\bibinfo {year}
  {2011})}\BibitemShut {NoStop}%
\bibitem [{\citenamefont {Fiaschi}\ \emph {et~al.}(2021)\citenamefont
  {Fiaschi}, \citenamefont {Hensen}, \citenamefont {Wallucks}, \citenamefont
  {Benevides}, \citenamefont {Li}, \citenamefont {Alegre},\ and\ \citenamefont
  {Gr\"oblacher}}]{fiaschi_natphot_optomechanical_quantum_teleportation}%
  \BibitemOpen
  \bibfield  {author} {\bibinfo {author} {\bibfnamefont {N.}~\bibnamefont
  {Fiaschi}}, \bibinfo {author} {\bibfnamefont {B.}~\bibnamefont {Hensen}},
  \bibinfo {author} {\bibfnamefont {A.}~\bibnamefont {Wallucks}}, \bibinfo
  {author} {\bibfnamefont {R.}~\bibnamefont {Benevides}}, \bibinfo {author}
  {\bibfnamefont {J.}~\bibnamefont {Li}}, \bibinfo {author} {\bibfnamefont
  {T.~P.~M.}\ \bibnamefont {Alegre}},\ and\ \bibinfo {author} {\bibfnamefont
  {S.}~\bibnamefont {Gr\"oblacher}},\ }\bibfield  {title} {\bibinfo {title}
  {Optomechanical quantum teleportation},\ }\href
  {https://doi.org/10.1038/s41566-021-00866-z} {\bibfield  {journal} {\bibinfo
  {journal} {Nat. Photonics}\ }\textbf {\bibinfo {volume} {15}},\ \bibinfo
  {pages} {817} (\bibinfo {year} {2021})}\BibitemShut {NoStop}%
\bibitem [{\citenamefont {Poot}\ \emph {et~al.}(2015)\citenamefont {Poot},
  \citenamefont {Fong},\ and\ \citenamefont {Tang}}]{poot_NJP_Yfeedback}%
  \BibitemOpen
  \bibfield  {author} {\bibinfo {author} {\bibfnamefont {M.}~\bibnamefont
  {Poot}}, \bibinfo {author} {\bibfnamefont {K.~Y.}\ \bibnamefont {Fong}},\
  and\ \bibinfo {author} {\bibfnamefont {H.~X.}\ \bibnamefont {Tang}},\
  }\bibfield  {title} {\bibinfo {title} {Deep feedback-stabilized parametric
  squeezing in an opto-electromechanical system},\ }\href
  {https://doi.org/10.1088/1367-2630/17/4/043056} {\bibfield  {journal}
  {\bibinfo  {journal} {New J. Phys.}\ }\textbf {\bibinfo {volume} {17}},\
  \bibinfo {pages} {043056} (\bibinfo {year} {2015})}\BibitemShut {NoStop}%
\bibitem [{\citenamefont {H{\o}j}\ \emph {et~al.}(2021)\citenamefont {H{\o}j},
  \citenamefont {Wang}, \citenamefont {Gao}, \citenamefont {Hoff},
  \citenamefont {Sigmund},\ and\ \citenamefont
  {Andersen}}]{Hoej_NatureCommunications_trampoline_inverse_design}%
  \BibitemOpen
  \bibfield  {author} {\bibinfo {author} {\bibfnamefont {D.}~\bibnamefont
  {H{\o}j}}, \bibinfo {author} {\bibfnamefont {F.}~\bibnamefont {Wang}},
  \bibinfo {author} {\bibfnamefont {W.}~\bibnamefont {Gao}}, \bibinfo {author}
  {\bibfnamefont {U.~B.}\ \bibnamefont {Hoff}}, \bibinfo {author}
  {\bibfnamefont {O.}~\bibnamefont {Sigmund}},\ and\ \bibinfo {author}
  {\bibfnamefont {U.~L.}\ \bibnamefont {Andersen}},\ }\bibfield  {title}
  {\bibinfo {title} {Ultra-coherent nanomechanical resonators based on inverse
  design},\ }\href {https://doi.org/10.1038/s41467-021-26102-4} {\bibfield
  {journal} {\bibinfo  {journal} {Nat. Commun.}\ }\textbf {\bibinfo {volume}
  {12}},\ \bibinfo {pages} {1} (\bibinfo {year} {2021})}\BibitemShut {NoStop}%
\bibitem [{\citenamefont {Unterreithmeier}\ \emph {et~al.}(2010)\citenamefont
  {Unterreithmeier}, \citenamefont {Faust},\ and\ \citenamefont
  {Kotthaus}}]{Unterreithmeier_PRL_damping}%
  \BibitemOpen
  \bibfield  {author} {\bibinfo {author} {\bibfnamefont {Q.~P.}\ \bibnamefont
  {Unterreithmeier}}, \bibinfo {author} {\bibfnamefont {T.}~\bibnamefont
  {Faust}},\ and\ \bibinfo {author} {\bibfnamefont {J.~P.}\ \bibnamefont
  {Kotthaus}},\ }\bibfield  {title} {\bibinfo {title} {Damping of
  nanomechanical resonators},\ }\href
  {https://doi.org/10.1103/physrevlett.105.027205} {\bibfield  {journal}
  {\bibinfo  {journal} {Phys. Rev. Lett.}\ }\textbf {\bibinfo {volume} {105}},\
  \bibinfo {pages} {027205} (\bibinfo {year} {2010})}\BibitemShut {NoStop}%
\bibitem [{\citenamefont {Norte}\ \emph {et~al.}(2016)\citenamefont {Norte},
  \citenamefont {Moura},\ and\ \citenamefont
  {Gr\"oblacher}}]{Norte_PhysicalReviewLetters_SiN_yield_stress}%
  \BibitemOpen
  \bibfield  {author} {\bibinfo {author} {\bibfnamefont {R.}~\bibnamefont
  {Norte}}, \bibinfo {author} {\bibfnamefont {J.}~\bibnamefont {Moura}},\ and\
  \bibinfo {author} {\bibfnamefont {S.}~\bibnamefont {Gr\"oblacher}},\
  }\bibfield  {title} {\bibinfo {title} {Mechanical resonators for quantum
  optomechanics experiments at room temperature},\ }\href
  {https://doi.org/10.1103/physrevlett.116.147202} {\bibfield  {journal}
  {\bibinfo  {journal} {Phys. Rev. Lett.}\ }\textbf {\bibinfo {volume} {116}},\
  \bibinfo {pages} {147202} (\bibinfo {year} {2016})}\BibitemShut {NoStop}%
\bibitem [{\citenamefont {Heinrich}\ \emph {et~al.}(2021)\citenamefont
  {Heinrich}, \citenamefont {Oliver}, \citenamefont {Vandersypen},
  \citenamefont {Ardavan}, \citenamefont {Sessoli}, \citenamefont {Loss},
  \citenamefont {Jayich}, \citenamefont {Fernandez-Rossier}, \citenamefont
  {Laucht},\ and\ \citenamefont
  {Morello}}]{Heinrich_NatureNanotechnology_quantum_coherence}%
  \BibitemOpen
  \bibfield  {author} {\bibinfo {author} {\bibfnamefont {A.~J.}\ \bibnamefont
  {Heinrich}}, \bibinfo {author} {\bibfnamefont {W.~D.}\ \bibnamefont
  {Oliver}}, \bibinfo {author} {\bibfnamefont {L.~M.~K.}\ \bibnamefont
  {Vandersypen}}, \bibinfo {author} {\bibfnamefont {A.}~\bibnamefont
  {Ardavan}}, \bibinfo {author} {\bibfnamefont {R.}~\bibnamefont {Sessoli}},
  \bibinfo {author} {\bibfnamefont {D.}~\bibnamefont {Loss}}, \bibinfo {author}
  {\bibfnamefont {A.~B.}\ \bibnamefont {Jayich}}, \bibinfo {author}
  {\bibfnamefont {J.}~\bibnamefont {Fernandez-Rossier}}, \bibinfo {author}
  {\bibfnamefont {A.}~\bibnamefont {Laucht}},\ and\ \bibinfo {author}
  {\bibfnamefont {A.}~\bibnamefont {Morello}},\ }\bibfield  {title} {\bibinfo
  {title} {Quantum-coherent nanoscience},\ }\href
  {https://doi.org/10.1038/s41565-021-00994-1} {\bibfield  {journal} {\bibinfo
  {journal} {Nature Nanotechnology}\ }\textbf {\bibinfo {volume} {16}},\
  \bibinfo {pages} {1318} (\bibinfo {year} {2021})}\BibitemShut {NoStop}%
\bibitem [{\citenamefont {Eichenfield}\ \emph {et~al.}(2009)\citenamefont
  {Eichenfield}, \citenamefont {Chan}, \citenamefont {Camacho}, \citenamefont
  {Vahala},\ and\ \citenamefont
  {Painter}}]{eichenfield_nature_optomechanical_crystals}%
  \BibitemOpen
  \bibfield  {author} {\bibinfo {author} {\bibfnamefont {M.}~\bibnamefont
  {Eichenfield}}, \bibinfo {author} {\bibfnamefont {J.}~\bibnamefont {Chan}},
  \bibinfo {author} {\bibfnamefont {R.~M.}\ \bibnamefont {Camacho}}, \bibinfo
  {author} {\bibfnamefont {K.~J.}\ \bibnamefont {Vahala}},\ and\ \bibinfo
  {author} {\bibfnamefont {O.}~\bibnamefont {Painter}},\ }\bibfield  {title}
  {\bibinfo {title} {Optomechanical crystals},\ }\href
  {http://dx.doi.org/10.1038/nature08524} {\bibfield  {journal} {\bibinfo
  {journal} {Nature}\ }\textbf {\bibinfo {volume} {462}},\ \bibinfo {pages}
  {78} (\bibinfo {year} {2009})}\BibitemShut {NoStop}%
\bibitem [{\citenamefont {Poot}\ and\ \citenamefont
  {Tang}(2014)}]{poot_apl_phaseshifter}%
  \BibitemOpen
  \bibfield  {author} {\bibinfo {author} {\bibfnamefont {M.}~\bibnamefont
  {Poot}}\ and\ \bibinfo {author} {\bibfnamefont {H.~X.}\ \bibnamefont
  {Tang}},\ }\bibfield  {title} {\bibinfo {title} {Broadband
  nanoelectromechanical phase shifting of light on a chip},\ }\href
  {https://doi.org/10.1063/1.4864257} {\bibfield  {journal} {\bibinfo
  {journal} {Appl. Phys. Lett.}\ }\textbf {\bibinfo {volume} {104}},\ \bibinfo
  {pages} {061101} (\bibinfo {year} {2014})}\BibitemShut {NoStop}%
\bibitem [{\citenamefont {Bagheri}\ \emph {et~al.}(2013)\citenamefont
  {Bagheri}, \citenamefont {Poot}, \citenamefont {Fan}, \citenamefont
  {Marquardt},\ and\ \citenamefont {Tang}}]{Bagheri_PRL_cavity_sync}%
  \BibitemOpen
  \bibfield  {author} {\bibinfo {author} {\bibfnamefont {M.}~\bibnamefont
  {Bagheri}}, \bibinfo {author} {\bibfnamefont {M.}~\bibnamefont {Poot}},
  \bibinfo {author} {\bibfnamefont {L.}~\bibnamefont {Fan}}, \bibinfo {author}
  {\bibfnamefont {F.}~\bibnamefont {Marquardt}},\ and\ \bibinfo {author}
  {\bibfnamefont {H.~X.}\ \bibnamefont {Tang}},\ }\bibfield  {title} {\bibinfo
  {title} {Photonic cavity synchronization of nanomechanical oscillators},\
  }\href {https://doi.org/10.1103/physrevlett.111.213902} {\bibfield  {journal}
  {\bibinfo  {journal} {Phys. Rev. Lett.}\ }\textbf {\bibinfo {volume} {111}},\
  \bibinfo {pages} {213902} (\bibinfo {year} {2013})}\BibitemShut {NoStop}%
\bibitem [{\citenamefont {Cole}\ \emph {et~al.}(2011)\citenamefont {Cole},
  \citenamefont {Wilson-Rae}, \citenamefont {Werbach}, \citenamefont {Vanner},\
  and\ \citenamefont {Aspelmeyer}}]{cole_natcomm_phonon_tunnelling}%
  \BibitemOpen
  \bibfield  {author} {\bibinfo {author} {\bibfnamefont {G.~D.}\ \bibnamefont
  {Cole}}, \bibinfo {author} {\bibfnamefont {I.}~\bibnamefont {Wilson-Rae}},
  \bibinfo {author} {\bibfnamefont {K.}~\bibnamefont {Werbach}}, \bibinfo
  {author} {\bibfnamefont {M.~R.}\ \bibnamefont {Vanner}},\ and\ \bibinfo
  {author} {\bibfnamefont {M.}~\bibnamefont {Aspelmeyer}},\ }\bibfield  {title}
  {\bibinfo {title} {Phonon-tunnelling dissipation in mechanical resonators},\
  }\href {https://doi.org/10.1038/ncomms1212} {\bibfield  {journal} {\bibinfo
  {journal} {Nature Communications}\ }\textbf {\bibinfo {volume} {2}},\
  \bibinfo {pages} {231} (\bibinfo {year} {2011})}\BibitemShut {NoStop}%
\bibitem [{\citenamefont {Fong}\ \emph {et~al.}(2019)\citenamefont {Fong},
  \citenamefont {Jin}, \citenamefont {Poot}, \citenamefont {Bruch},\ and\
  \citenamefont {Tang}}]{Fong_NanoLetters_phonon_coupling}%
  \BibitemOpen
  \bibfield  {author} {\bibinfo {author} {\bibfnamefont {K.~Y.}\ \bibnamefont
  {Fong}}, \bibinfo {author} {\bibfnamefont {D.}~\bibnamefont {Jin}}, \bibinfo
  {author} {\bibfnamefont {M.}~\bibnamefont {Poot}}, \bibinfo {author}
  {\bibfnamefont {A.}~\bibnamefont {Bruch}},\ and\ \bibinfo {author}
  {\bibfnamefont {H.~X.}\ \bibnamefont {Tang}},\ }\bibfield  {title} {\bibinfo
  {title} {Phonon coupling between a nanomechanical resonator and a quantum
  fluid},\ }\href {https://doi.org/10.1021/acs.nanolett.9b00821} {\bibfield
  {journal} {\bibinfo  {journal} {Nano Lett.}\ }\textbf {\bibinfo {volume}
  {19}},\ \bibinfo {pages} {3716} (\bibinfo {year} {2019})}\BibitemShut
  {NoStop}%
\bibitem [{\citenamefont {Bereyhi}\ \emph {et~al.}(2021)\citenamefont
  {Bereyhi}, \citenamefont {Arabmoheghi}, \citenamefont {Fedorov},
  \citenamefont {Beccari}, \citenamefont {Huang}, \citenamefont {Kippenberg},\
  and\ \citenamefont {Engelsen}}]{Bereyhi_arXiv_ultrahighQ}%
  \BibitemOpen
  \bibfield  {author} {\bibinfo {author} {\bibfnamefont {M.~J.}\ \bibnamefont
  {Bereyhi}}, \bibinfo {author} {\bibfnamefont {A.}~\bibnamefont
  {Arabmoheghi}}, \bibinfo {author} {\bibfnamefont {S.~A.}\ \bibnamefont
  {Fedorov}}, \bibinfo {author} {\bibfnamefont {A.}~\bibnamefont {Beccari}},
  \bibinfo {author} {\bibfnamefont {G.}~\bibnamefont {Huang}}, \bibinfo
  {author} {\bibfnamefont {T.~J.}\ \bibnamefont {Kippenberg}},\ and\ \bibinfo
  {author} {\bibfnamefont {N.~J.}\ \bibnamefont {Engelsen}},\ }\bibfield
  {title} {\bibinfo {title} {Nanomechanical resonators with ultra-high-$q$
  perimeter modes},\ }\href {https://arxiv.org/abs/2108.03615} {\bibfield
  {journal} {\bibinfo  {journal} {arXiv:2108.03615v2}\ } (\bibinfo {year}
  {2021})},\ \bibinfo {note} {https://arxiv.org/abs/2108.03615. (accessed March
  14, 2022)},\ \Eprint {https://arxiv.org/abs/2108.03615} {arXiv:2108.03615
  [physics.app-ph]} \BibitemShut {NoStop}%
\bibitem [{\citenamefont {Ghadimi}\ \emph {et~al.}(2018)\citenamefont
  {Ghadimi}, \citenamefont {Fedorov}, \citenamefont {Engelsen}, \citenamefont
  {Bereyhi}, \citenamefont {Schilling}, \citenamefont {Wilson},\ and\
  \citenamefont {Kippenberg}}]{Ghadimi_Science_strain_engineering}%
  \BibitemOpen
  \bibfield  {author} {\bibinfo {author} {\bibfnamefont {A.~H.}\ \bibnamefont
  {Ghadimi}}, \bibinfo {author} {\bibfnamefont {S.~A.}\ \bibnamefont
  {Fedorov}}, \bibinfo {author} {\bibfnamefont {N.~J.}\ \bibnamefont
  {Engelsen}}, \bibinfo {author} {\bibfnamefont {M.~J.}\ \bibnamefont
  {Bereyhi}}, \bibinfo {author} {\bibfnamefont {R.}~\bibnamefont {Schilling}},
  \bibinfo {author} {\bibfnamefont {D.~J.}\ \bibnamefont {Wilson}},\ and\
  \bibinfo {author} {\bibfnamefont {T.~J.}\ \bibnamefont {Kippenberg}},\
  }\bibfield  {title} {\bibinfo {title} {Elastic strain engineering for
  ultralow mechanical dissipation},\ }\href
  {https://doi.org/10.1126/science.aar6939} {\bibfield  {journal} {\bibinfo
  {journal} {Science}\ }\textbf {\bibinfo {volume} {360}},\ \bibinfo {pages}
  {764} (\bibinfo {year} {2018})}\BibitemShut {NoStop}%
\bibitem [{\citenamefont {Beccari}\ \emph {et~al.}(2022)\citenamefont
  {Beccari}, \citenamefont {Visani}, \citenamefont {Fedorov}, \citenamefont
  {Bereyhi}, \citenamefont {Boureau}, \citenamefont {Engelsen},\ and\
  \citenamefont {Kippenberg}}]{Beccari2022_nature_10B_quality}%
  \BibitemOpen
  \bibfield  {author} {\bibinfo {author} {\bibfnamefont {A.}~\bibnamefont
  {Beccari}}, \bibinfo {author} {\bibfnamefont {D.~A.}\ \bibnamefont {Visani}},
  \bibinfo {author} {\bibfnamefont {S.~A.}\ \bibnamefont {Fedorov}}, \bibinfo
  {author} {\bibfnamefont {M.~J.}\ \bibnamefont {Bereyhi}}, \bibinfo {author}
  {\bibfnamefont {V.}~\bibnamefont {Boureau}}, \bibinfo {author} {\bibfnamefont
  {N.~J.}\ \bibnamefont {Engelsen}},\ and\ \bibinfo {author} {\bibfnamefont
  {T.~J.}\ \bibnamefont {Kippenberg}},\ }\bibfield  {title} {\bibinfo {title}
  {Strained crystalline nanomechanical resonators with quality factors above 10
  billion},\ }\bibfield  {journal} {\bibinfo  {journal} {Nat. Phys.}\ }\href
  {https://doi.org/10.1038/s41567-021-01498-4} {10.1038/s41567-021-01498-4}
  (\bibinfo {year} {2022})\BibitemShut {NoStop}%
\bibitem [{\citenamefont {Hoch}\ \emph {et~al.}(2022)\citenamefont {Hoch},
  \citenamefont {Yao},\ and\ \citenamefont {Poot}}]{Hoch_Sbeam}%
  \BibitemOpen
  \bibfield  {author} {\bibinfo {author} {\bibfnamefont {D.}~\bibnamefont
  {Hoch}}, \bibinfo {author} {\bibfnamefont {X.}~\bibnamefont {Yao}},\ and\
  \bibinfo {author} {\bibfnamefont {M.}~\bibnamefont {Poot}},\ }\bibfield
  {title} {\bibinfo {title} {Geometric tuning of stress in silicon nitride beam
  resonators},\ }\href@noop {} {\bibfield  {journal} {\bibinfo  {journal} {in
  preparation}\ } (\bibinfo {year} {2022})}\BibitemShut {NoStop}%
\bibitem [{\citenamefont {Schmid}\ \emph {et~al.}(2011)\citenamefont {Schmid},
  \citenamefont {Jensen}, \citenamefont {Nielsen},\ and\ \citenamefont
  {Boisen}}]{Schmid_PhysRevB_damping_model}%
  \BibitemOpen
  \bibfield  {author} {\bibinfo {author} {\bibfnamefont {S.}~\bibnamefont
  {Schmid}}, \bibinfo {author} {\bibfnamefont {K.~D.}\ \bibnamefont {Jensen}},
  \bibinfo {author} {\bibfnamefont {K.~H.}\ \bibnamefont {Nielsen}},\ and\
  \bibinfo {author} {\bibfnamefont {A.}~\bibnamefont {Boisen}},\ }\bibfield
  {title} {\bibinfo {title} {Damping mechanisms in high-{Q} micro and
  nanomechanical string resonators},\ }\href
  {https://doi.org/10.1103/physrevb.84.165307} {\bibfield  {journal} {\bibinfo
  {journal} {Phys. Rev. B}\ }\textbf {\bibinfo {volume} {84}},\ \bibinfo
  {pages} {165307} (\bibinfo {year} {2011})}\BibitemShut {NoStop}%
\bibitem [{\citenamefont {Hoch}\ \emph {et~al.}(2021)\citenamefont {Hoch},
  \citenamefont {Haas}, \citenamefont {Moller}, \citenamefont {Sommer},
  \citenamefont {Soubelet}, \citenamefont {Finley},\ and\ \citenamefont
  {Poot}}]{hoch_MM_mode_mapping}%
  \BibitemOpen
  \bibfield  {author} {\bibinfo {author} {\bibfnamefont {D.}~\bibnamefont
  {Hoch}}, \bibinfo {author} {\bibfnamefont {K.-J.}\ \bibnamefont {Haas}},
  \bibinfo {author} {\bibfnamefont {L.}~\bibnamefont {Moller}}, \bibinfo
  {author} {\bibfnamefont {T.}~\bibnamefont {Sommer}}, \bibinfo {author}
  {\bibfnamefont {P.}~\bibnamefont {Soubelet}}, \bibinfo {author}
  {\bibfnamefont {J.~J.}\ \bibnamefont {Finley}},\ and\ \bibinfo {author}
  {\bibfnamefont {M.}~\bibnamefont {Poot}},\ }\bibfield  {title} {\bibinfo
  {title} {Efficient optomechanical mode-shape mapping of micromechanical
  devices},\ }\href {https://doi.org/10.3390/mi12080880} {\bibfield  {journal}
  {\bibinfo  {journal} {Micromachines}\ }\textbf {\bibinfo {volume} {12}},\
  \bibinfo {pages} {880} (\bibinfo {year} {2021})}\BibitemShut {NoStop}%
\bibitem [{\citenamefont {Etaki}\ \emph {et~al.}(2008)\citenamefont {Etaki},
  \citenamefont {Poot}, \citenamefont {Mahboob}, \citenamefont {Onomitsu},
  \citenamefont {Yamaguchi},\ and\ \citenamefont {van~der
  Zant}}]{etaki_natphys_squid}%
  \BibitemOpen
  \bibfield  {author} {\bibinfo {author} {\bibfnamefont {S.}~\bibnamefont
  {Etaki}}, \bibinfo {author} {\bibfnamefont {M.}~\bibnamefont {Poot}},
  \bibinfo {author} {\bibfnamefont {I.}~\bibnamefont {Mahboob}}, \bibinfo
  {author} {\bibfnamefont {K.}~\bibnamefont {Onomitsu}}, \bibinfo {author}
  {\bibfnamefont {H.}~\bibnamefont {Yamaguchi}},\ and\ \bibinfo {author}
  {\bibfnamefont {H.~S.~J.}\ \bibnamefont {van~der Zant}},\ }\bibfield  {title}
  {\bibinfo {title} {Motion detection of a micromechanical resonator embedded
  in a d.c. squid},\ }\href {https://doi.org/10.1038/nphys1057} {\bibfield
  {journal} {\bibinfo  {journal} {Nat Phys}\ }\textbf {\bibinfo {volume} {4}},\
  \bibinfo {pages} {785} (\bibinfo {year} {2008})}\BibitemShut {NoStop}%
\bibitem [{\citenamefont {Erbil}\ \emph {et~al.}(2020)\citenamefont {Erbil},
  \citenamefont {Hatipoglu}, \citenamefont {Yanik}, \citenamefont {Ghavami},
  \citenamefont {Ari}, \citenamefont {Yuksel},\ and\ \citenamefont
  {Hanay}}]{erbil_PRL_buckling_MEMS}%
  \BibitemOpen
  \bibfield  {author} {\bibinfo {author} {\bibfnamefont {S.~O.}\ \bibnamefont
  {Erbil}}, \bibinfo {author} {\bibfnamefont {U.}~\bibnamefont {Hatipoglu}},
  \bibinfo {author} {\bibfnamefont {C.}~\bibnamefont {Yanik}}, \bibinfo
  {author} {\bibfnamefont {M.}~\bibnamefont {Ghavami}}, \bibinfo {author}
  {\bibfnamefont {A.~B.}\ \bibnamefont {Ari}}, \bibinfo {author} {\bibfnamefont
  {M.}~\bibnamefont {Yuksel}},\ and\ \bibinfo {author} {\bibfnamefont {M.~S.}\
  \bibnamefont {Hanay}},\ }\bibfield  {title} {\bibinfo {title} {Full
  electrostatic control of nanomechanical buckling},\ }\href
  {https://doi.org/10.1103/PhysRevLett.124.046101} {\bibfield  {journal}
  {\bibinfo  {journal} {Phys. Rev. Lett.}\ }\textbf {\bibinfo {volume} {124}},\
  \bibinfo {pages} {046101} (\bibinfo {year} {2020})}\BibitemShut {NoStop}%
\bibitem [{\citenamefont {Nayfeh}\ \emph {et~al.}(1995)\citenamefont {Nayfeh},
  \citenamefont {Kreider},\ and\ \citenamefont {Anderson}}]{nayfeh_buckled}%
  \BibitemOpen
  \bibfield  {author} {\bibinfo {author} {\bibfnamefont {A.~H.}\ \bibnamefont
  {Nayfeh}}, \bibinfo {author} {\bibfnamefont {W.}~\bibnamefont {Kreider}},\
  and\ \bibinfo {author} {\bibfnamefont {T.~J.}\ \bibnamefont {Anderson}},\
  }\bibfield  {title} {\bibinfo {title} {Investigation of natural frequencies
  and mode shapes of bukcled beams},\ }\href {https://doi.org/10.2514/3.12669}
  {\bibfield  {journal} {\bibinfo  {journal} {AIAA Journal}\ }\textbf {\bibinfo
  {volume} {33}},\ \bibinfo {pages} {1121} (\bibinfo {year}
  {1995})}\BibitemShut {NoStop}%
\bibitem [{\citenamefont {Charlot}\ \emph {et~al.}(2008)\citenamefont
  {Charlot}, \citenamefont {Sun}, \citenamefont {Yamashita}, \citenamefont
  {Fujita},\ and\ \citenamefont {Toshiyoshi}}]{charlot_JMM_buckling_memory}%
  \BibitemOpen
  \bibfield  {author} {\bibinfo {author} {\bibfnamefont {B.}~\bibnamefont
  {Charlot}}, \bibinfo {author} {\bibfnamefont {W.}~\bibnamefont {Sun}},
  \bibinfo {author} {\bibfnamefont {K.}~\bibnamefont {Yamashita}}, \bibinfo
  {author} {\bibfnamefont {H.}~\bibnamefont {Fujita}},\ and\ \bibinfo {author}
  {\bibfnamefont {H.}~\bibnamefont {Toshiyoshi}},\ }\bibfield  {title}
  {\bibinfo {title} {Bistable nanowire for micromechanical memory},\ }\href
  {https://doi.org/10.1088/0960-1317/18/4/045005} {\bibfield  {journal}
  {\bibinfo  {journal} {J Micromechanics Microengineering}\ }\textbf {\bibinfo
  {volume} {18}},\ \bibinfo {pages} {045005} (\bibinfo {year}
  {2008})}\BibitemShut {NoStop}%
\bibitem [{\citenamefont {Bagheri}\ \emph {et~al.}(2011)\citenamefont
  {Bagheri}, \citenamefont {Poot}, \citenamefont {Li}, \citenamefont
  {Pernice},\ and\ \citenamefont {Tang}}]{bagheri_natnano_high_amplitude}%
  \BibitemOpen
  \bibfield  {author} {\bibinfo {author} {\bibfnamefont {M.}~\bibnamefont
  {Bagheri}}, \bibinfo {author} {\bibfnamefont {M.}~\bibnamefont {Poot}},
  \bibinfo {author} {\bibfnamefont {M.}~\bibnamefont {Li}}, \bibinfo {author}
  {\bibfnamefont {W.~P.~H.}\ \bibnamefont {Pernice}},\ and\ \bibinfo {author}
  {\bibfnamefont {H.~X.}\ \bibnamefont {Tang}},\ }\bibfield  {title} {\bibinfo
  {title} {Dynamic manipulation of nanomechanical resonators in the
  high-amplitude regime and non-volatile mechanical memory operation},\ }\href
  {https://doi.org/10.1038/nnano.2011.180} {\bibfield  {journal} {\bibinfo
  {journal} {Nat Nano}\ }\textbf {\bibinfo {volume} {6}},\ \bibinfo {pages}
  {726} (\bibinfo {year} {2011})}\BibitemShut {NoStop}%
\bibitem [{\citenamefont {Kim}\ \emph {et~al.}(2021)\citenamefont {Kim},
  \citenamefont {Bunyan}, \citenamefont {Ferrari}, \citenamefont {Kanj},
  \citenamefont {Vakakis}, \citenamefont {van~der Zande},\ and\ \citenamefont
  {Tawfick}}]{kim_NL_buckling_SiN_drum_array}%
  \BibitemOpen
  \bibfield  {author} {\bibinfo {author} {\bibfnamefont {S.}~\bibnamefont
  {Kim}}, \bibinfo {author} {\bibfnamefont {J.}~\bibnamefont {Bunyan}},
  \bibinfo {author} {\bibfnamefont {P.~F.}\ \bibnamefont {Ferrari}}, \bibinfo
  {author} {\bibfnamefont {A.}~\bibnamefont {Kanj}}, \bibinfo {author}
  {\bibfnamefont {A.~F.}\ \bibnamefont {Vakakis}}, \bibinfo {author}
  {\bibfnamefont {A.~M.}\ \bibnamefont {van~der Zande}},\ and\ \bibinfo
  {author} {\bibfnamefont {S.}~\bibnamefont {Tawfick}},\ }\bibfield  {title}
  {\bibinfo {title} {Buckling-mediated phase transitions in
  nano-electromechanical phononic waveguides},\ }\href
  {https://doi.org/10.1021/acs.nanolett.1c00764} {\bibfield  {journal}
  {\bibinfo  {journal} {Nano Lett.}\ }\textbf {\bibinfo {volume} {21}},\
  \bibinfo {pages} {6416} (\bibinfo {year} {2021})},\ \bibinfo {note} {pMID:
  34320324},\ \Eprint
  {https://arxiv.org/abs/https://doi.org/10.1021/acs.nanolett.1c00764}
  {https://doi.org/10.1021/acs.nanolett.1c00764} \BibitemShut {NoStop}%
\bibitem [{\citenamefont {Poot}\ and\ \citenamefont {van~der
  Zant}(2012)}]{poot_physrep_quantum_regime}%
  \BibitemOpen
  \bibfield  {author} {\bibinfo {author} {\bibfnamefont {M.}~\bibnamefont
  {Poot}}\ and\ \bibinfo {author} {\bibfnamefont {H.~S.}\ \bibnamefont {van~der
  Zant}},\ }\bibfield  {title} {\bibinfo {title} {Mechanical systems in the
  quantum regime},\ }\href {https://doi.org/10.1016/j.physrep.2011.12.004}
  {\bibfield  {journal} {\bibinfo  {journal} {Phys. Rep.}\ }\textbf {\bibinfo
  {volume} {511}},\ \bibinfo {pages} {273} (\bibinfo {year}
  {2012})}\BibitemShut {NoStop}%
\bibitem [{\citenamefont {Cleland}(2003)}]{cleland_nanomechanics}%
  \BibitemOpen
  \bibfield  {author} {\bibinfo {author} {\bibfnamefont {A.}~\bibnamefont
  {Cleland}},\ }\href@noop {} {\emph {\bibinfo {title} {Foundations of
  Nanomechanics}}}\ (\bibinfo  {publisher} {Springer},\ \bibinfo {year}
  {2003})\BibitemShut {NoStop}%
\bibitem [{\citenamefont {Landau}\ and\ \citenamefont
  {Lifshitz}(1986)}]{LL_elasticity}%
  \BibitemOpen
  \bibfield  {author} {\bibinfo {author} {\bibfnamefont {L.~D.}\ \bibnamefont
  {Landau}}\ and\ \bibinfo {author} {\bibfnamefont {E.~M.}\ \bibnamefont
  {Lifshitz}},\ }\href@noop {} {\emph {\bibinfo {title} {Theory of
  elasticity}}}\ (\bibinfo  {publisher} {Butterworth-Heineman},\ \bibinfo
  {year} {1986})\BibitemShut {NoStop}%
\bibitem [{\citenamefont {Wells}(1967)}]{Wells_Lagrangian_dynamics}%
  \BibitemOpen
  \bibfield  {author} {\bibinfo {author} {\bibfnamefont {D.~A.}\ \bibnamefont
  {Wells}},\ }\href@noop {} {\emph {\bibinfo {title} {Schaum's Outline of
  Theory and Problems of Lagrangian Dynamics}}}\ (\bibinfo  {publisher}
  {McGraw-Hill},\ \bibinfo {year} {1967})\BibitemShut {NoStop}%
\bibitem [{\citenamefont {Poot}\ \emph {et~al.}(2007)\citenamefont {Poot},
  \citenamefont {Witkamp}, \citenamefont {Otte},\ and\ \citenamefont {van~der
  Zant}}]{poot_PSSB_modelling_CNT}%
  \BibitemOpen
  \bibfield  {author} {\bibinfo {author} {\bibfnamefont {M.}~\bibnamefont
  {Poot}}, \bibinfo {author} {\bibfnamefont {B.}~\bibnamefont {Witkamp}},
  \bibinfo {author} {\bibfnamefont {M.~A.}\ \bibnamefont {Otte}},\ and\
  \bibinfo {author} {\bibfnamefont {H.~S.~J.}\ \bibnamefont {van~der Zant}},\
  }\bibfield  {title} {\bibinfo {title} {Modelling suspended carbon nanotube
  resonators},\ }\href {http://dx.doi.org/10.1002/pssb.200776130} {\bibfield
  {journal} {\bibinfo  {journal} {Phys. Stat. Sol. (b)}\ }\textbf {\bibinfo
  {volume} {244}},\ \bibinfo {pages} {4252} (\bibinfo {year}
  {2007})}\BibitemShut {NoStop}%
\bibitem [{\citenamefont {Westra}\ \emph {et~al.}(2010)\citenamefont {Westra},
  \citenamefont {Poot}, \citenamefont {van~der Zant},\ and\ \citenamefont
  {Venstra}}]{westra_PRL_coupled}%
  \BibitemOpen
  \bibfield  {author} {\bibinfo {author} {\bibfnamefont {H.~J.~R.}\
  \bibnamefont {Westra}}, \bibinfo {author} {\bibfnamefont {M.}~\bibnamefont
  {Poot}}, \bibinfo {author} {\bibfnamefont {H.~S.~J.}\ \bibnamefont {van~der
  Zant}},\ and\ \bibinfo {author} {\bibfnamefont {W.~J.}\ \bibnamefont
  {Venstra}},\ }\bibfield  {title} {\bibinfo {title} {Nonlinear modal
  interactions in clamped-clamped mechanical resonators},\ }\href
  {https://doi.org/10.1103/PhysRevLett.105.117205} {\bibfield  {journal}
  {\bibinfo  {journal} {Phys. Rev. Lett.}\ }\textbf {\bibinfo {volume} {105}},\
  \bibinfo {pages} {117205} (\bibinfo {year} {2010})}\BibitemShut {NoStop}%
\bibitem [{\citenamefont {Nayfeh}\ and\ \citenamefont
  {Mook}(1979)}]{nayfeh_nonlinear}%
  \BibitemOpen
  \bibfield  {author} {\bibinfo {author} {\bibfnamefont {A.~H.}\ \bibnamefont
  {Nayfeh}}\ and\ \bibinfo {author} {\bibfnamefont {D.~T.}\ \bibnamefont
  {Mook}},\ }\href@noop {} {\emph {\bibinfo {title} {Nonlinear Oscillations}}}\
  (\bibinfo  {publisher} {Wiley},\ \bibinfo {year} {1979})\BibitemShut
  {NoStop}%
\bibitem [{Note1()}]{Note1}%
  \BibitemOpen
  \bibinfo {note} {Note that due to the symmetry with respect to $V=0$ (see
  Sec.~\ref {ssec:landscape}), there is no linear coupling ($\protect
  \ensuremath {\partial ^2E_\protect \mathrm {tot}/\partial U\partial V} = 0$
  for $V=0$) between $U$ and $V$ and the eigenmodes are purely y and z
  polarized.}\BibitemShut {Stop}%
\bibitem [{\citenamefont {Verbridge}\ \emph {et~al.}(2006)\citenamefont
  {Verbridge}, \citenamefont {Parpia}, \citenamefont {Reichenbach},
  \citenamefont {Bellan},\ and\ \citenamefont
  {Craighead}}]{Verbridge_JournalofAppliedPhysics_highstress_SiN}%
  \BibitemOpen
  \bibfield  {author} {\bibinfo {author} {\bibfnamefont {S.~S.}\ \bibnamefont
  {Verbridge}}, \bibinfo {author} {\bibfnamefont {J.~M.}\ \bibnamefont
  {Parpia}}, \bibinfo {author} {\bibfnamefont {R.~B.}\ \bibnamefont
  {Reichenbach}}, \bibinfo {author} {\bibfnamefont {L.~M.}\ \bibnamefont
  {Bellan}},\ and\ \bibinfo {author} {\bibfnamefont {H.~G.}\ \bibnamefont
  {Craighead}},\ }\bibfield  {title} {\bibinfo {title} {High quality factor
  resonance at room temperature with nanostrings under high tensile stress},\
  }\href {https://doi.org/10.1063/1.2204829} {\bibfield  {journal} {\bibinfo
  {journal} {J. Appl. Phys.}\ }\textbf {\bibinfo {volume} {99}},\ \bibinfo
  {pages} {124304} (\bibinfo {year} {2006})}\BibitemShut {NoStop}%
\bibitem [{\citenamefont {Ghadimi}\ \emph {et~al.}(2017)\citenamefont
  {Ghadimi}, \citenamefont {Wilson},\ and\ \citenamefont
  {Kippenberg}}]{Ghadimi_NanoLett_SiN_loss_engineering}%
  \BibitemOpen
  \bibfield  {author} {\bibinfo {author} {\bibfnamefont {A.~H.}\ \bibnamefont
  {Ghadimi}}, \bibinfo {author} {\bibfnamefont {D.~J.}\ \bibnamefont
  {Wilson}},\ and\ \bibinfo {author} {\bibfnamefont {T.~J.}\ \bibnamefont
  {Kippenberg}},\ }\bibfield  {title} {\bibinfo {title} {Radiation and internal
  loss engineering of high-stress silicon nitride nanobeams},\ }\href
  {https://doi.org/10.1021/acs.nanolett.7b00573} {\bibfield  {journal}
  {\bibinfo  {journal} {Nano Lett.}\ }\textbf {\bibinfo {volume} {17}},\
  \bibinfo {pages} {3501} (\bibinfo {year} {2017})}\BibitemShut {NoStop}%
\bibitem [{\citenamefont {Terrasanta}\ \emph {et~al.}(2022)\citenamefont
  {Terrasanta}, \citenamefont {Sommer}, \citenamefont {M\"{u}ller},
  \citenamefont {Althammer}, \citenamefont {Gross},\ and\ \citenamefont
  {Poot}}]{Terrasanta_OE_AlN_on_SiN}%
  \BibitemOpen
  \bibfield  {author} {\bibinfo {author} {\bibfnamefont {G.}~\bibnamefont
  {Terrasanta}}, \bibinfo {author} {\bibfnamefont {T.}~\bibnamefont {Sommer}},
  \bibinfo {author} {\bibfnamefont {M.}~\bibnamefont {M\"{u}ller}}, \bibinfo
  {author} {\bibfnamefont {M.}~\bibnamefont {Althammer}}, \bibinfo {author}
  {\bibfnamefont {R.}~\bibnamefont {Gross}},\ and\ \bibinfo {author}
  {\bibfnamefont {M.}~\bibnamefont {Poot}},\ }\bibfield  {title} {\bibinfo
  {title} {Aluminum nitride integration on silicon nitride photonic circuits: a
  hybrid approach towards on-chip nonlinear optics},\ }\href
  {https://doi.org/10.1364/OE.445465} {\bibfield  {journal} {\bibinfo
  {journal} {Opt. Express}\ }\textbf {\bibinfo {volume} {30}},\ \bibinfo
  {pages} {8537} (\bibinfo {year} {2022})}\BibitemShut {NoStop}%
\bibitem [{\citenamefont {B{\"u}ckle}\ \emph {et~al.}(2021)\citenamefont
  {B{\"u}ckle}, \citenamefont {Kla{\ss}}, \citenamefont {N{\"a}gele},
  \citenamefont {Braive},\ and\ \citenamefont
  {Weig}}]{buckle_PRAppl_tension_universal_length_dependence}%
  \BibitemOpen
  \bibfield  {author} {\bibinfo {author} {\bibfnamefont {M.}~\bibnamefont
  {B{\"u}ckle}}, \bibinfo {author} {\bibfnamefont {Y.~S.}\ \bibnamefont
  {Kla{\ss}}}, \bibinfo {author} {\bibfnamefont {F.~B.}\ \bibnamefont
  {N{\"a}gele}}, \bibinfo {author} {\bibfnamefont {R.}~\bibnamefont {Braive}},\
  and\ \bibinfo {author} {\bibfnamefont {E.~M.}\ \bibnamefont {Weig}},\
  }\bibfield  {title} {\bibinfo {title} {Universal length dependence of tensile
  stress in nanomechanical string resonators},\ }\href
  {https://doi.org/10.1103/PhysRevApplied.15.034063} {\bibfield  {journal}
  {\bibinfo  {journal} {Phys. Rev. Applied}\ }\textbf {\bibinfo {volume}
  {15}},\ \bibinfo {pages} {034063} (\bibinfo {year} {2021})}\BibitemShut
  {NoStop}%
\bibitem [{\citenamefont {Babaei~Gavan}\ \emph {et~al.}(2009)\citenamefont
  {Babaei~Gavan}, \citenamefont {van~der Drift}, \citenamefont {Venstra},
  \citenamefont {Zuiddam},\ and\ \citenamefont {van~der Zant H.
  S.~J.}}]{babei_gavan_JMM_undercut}%
  \BibitemOpen
  \bibfield  {author} {\bibinfo {author} {\bibfnamefont {K.}~\bibnamefont
  {Babaei~Gavan}}, \bibinfo {author} {\bibfnamefont {E.~W. J.~M.}\ \bibnamefont
  {van~der Drift}}, \bibinfo {author} {\bibfnamefont {W.~J.}\ \bibnamefont
  {Venstra}}, \bibinfo {author} {\bibfnamefont {M.~R.}\ \bibnamefont
  {Zuiddam}},\ and\ \bibinfo {author} {\bibnamefont {van~der Zant H. S.~J.}},\
  }\bibfield  {title} {\bibinfo {title} {Effect of undercut on the resonant
  behaviour of silicon nitride cantilevers},\ }\href
  {http://stacks.iop.org/0960-1317/19/i=3/a=035003} {\bibfield  {journal}
  {\bibinfo  {journal} {J. Micromechanics Microengineering}\ }\textbf {\bibinfo
  {volume} {19}},\ \bibinfo {pages} {035003} (\bibinfo {year}
  {2009})}\BibitemShut {NoStop}%
\bibitem [{Note2()}]{Note2}%
  \BibitemOpen
  \bibinfo {note} {The Cartesian coordinates used here correspond to the
  original, undeformed geometry.}\BibitemShut {Stop}%
\bibitem [{\citenamefont {Bereyhi}\ \emph {et~al.}(2019)\citenamefont
  {Bereyhi}, \citenamefont {Beccari}, \citenamefont {Fedorov}, \citenamefont
  {Ghadimi}, \citenamefont {Schilling}, \citenamefont {Wilson}, \citenamefont
  {Engelsen},\ and\ \citenamefont
  {Kippenberg}}]{Bereyhi_NanoLett_clamp_tapering}%
  \BibitemOpen
  \bibfield  {author} {\bibinfo {author} {\bibfnamefont {M.~J.}\ \bibnamefont
  {Bereyhi}}, \bibinfo {author} {\bibfnamefont {A.}~\bibnamefont {Beccari}},
  \bibinfo {author} {\bibfnamefont {S.~A.}\ \bibnamefont {Fedorov}}, \bibinfo
  {author} {\bibfnamefont {A.~H.}\ \bibnamefont {Ghadimi}}, \bibinfo {author}
  {\bibfnamefont {R.}~\bibnamefont {Schilling}}, \bibinfo {author}
  {\bibfnamefont {D.~J.}\ \bibnamefont {Wilson}}, \bibinfo {author}
  {\bibfnamefont {N.~J.}\ \bibnamefont {Engelsen}},\ and\ \bibinfo {author}
  {\bibfnamefont {T.~J.}\ \bibnamefont {Kippenberg}},\ }\bibfield  {title}
  {\bibinfo {title} {Clamp-tapering increases the quality factor of stressed
  nanobeams},\ }\href {https://doi.org/10.1021/acs.nanolett.8b04942} {\bibfield
   {journal} {\bibinfo  {journal} {Nano Lett.}\ }\textbf {\bibinfo {volume}
  {19}},\ \bibinfo {pages} {2329} (\bibinfo {year} {2019})}\BibitemShut
  {NoStop}%
\bibitem [{\citenamefont {Flensberg}(2006)}]{flensberg_NJP_electronphonon}%
  \BibitemOpen
  \bibfield  {author} {\bibinfo {author} {\bibfnamefont {K.}~\bibnamefont
  {Flensberg}},\ }\bibfield  {title} {\bibinfo {title} {Electron-vibron
  coupling in suspended nanotubes},\ }\href
  {http://stacks.iop.org/1367-2630/8/5} {\bibfield  {journal} {\bibinfo
  {journal} {New J. Phys.}\ }\textbf {\bibinfo {volume} {8}},\ \bibinfo {pages}
  {5} (\bibinfo {year} {2006})}\BibitemShut {NoStop}%
\bibitem [{\citenamefont {{COMSOL material library}}()}]{COMSOL_SiN}%
  \BibitemOpen
  \bibfield  {author} {\bibinfo {author} {\bibnamefont {{COMSOL material
  library}}},\ }\href@noop {} {\bibinfo {title} {{Si3N4 - Silicon nitride}}},\
  \bibinfo {howpublished} {{COMSOL Multiphysics v5.6}}\BibitemShut {NoStop}%
\end{thebibliography}%

\end{document}